\documentclass[12pt,a4paper]{article} 
\usepackage[latin1]{inputenc}% Para poder digitar os acentos da maneira usual
\usepackage[T1]{fontenc}
\usepackage[mathscr]{eucal}  % inclui a fonte matemática caligráfica $\mathscr{C}$
\usepackage{times}           % Para que o documento aceite a fonte Times Roman
\usepackage{enumerate}
\usepackage[dvips]{graphicx}
\usepackage{fancyheadings}
\usepackage{epsfig}
\usepackage{verbatim}
\usepackage{amsmath}

\newcommand{\bd}{\begin{displaymath}}
\newcommand{\ed}{\end{displaymath}}
\newcommand{\beq}{\begin{equation}}
\newcommand{\eeq}{\end{equation}}
\newcommand{\ba}{\begin{array}}
\newcommand{\ea}{\end{array}}
\newcommand{\bal}{\begin{align}}
\newcommand{\eal}{\end{align}}
\newcommand{\bpm}{\begin{pmatrix}}
\newcommand{\epm}{\end{pmatrix}}

\begin{document}

\title{The climate version of the Eta regional forecast model. 2.Evaluation of  the Eta CCS 
model performance against reanalysis data and surface observations.}

\author{I. A. Pisnichenko\footnote{Centro de Ciencia de Sistema Terrestre/Instituto Nacional 
de Pesquisas Espaciais, Cachoeira Paulista, SP, Brazil, (igor.pisnitchenko@cptec.inpe.br).Additional affiliation: A.M. Obukhov 
Institute of Atmospheric Physics, Russian Academy of Sciences, Moscow, Russia.}  
\and T.A. Tarasova\footnotemark[1] }

\date{ 10 December  2008}
\maketitle
\centerline{\bf Abstract}

    The climate version Eta CCS, prepared from the NCEP Eta forecast model, was
integrated over South America for the period from January 1979 to December 1983.
The model was driven by the two sets of boundary conditions derived from the
reanalysis and outputs of HadAM3P atmospheric global model. The  mean output
fields of precipitation, precipitation frequency, and near surface air
temperature, simulated by the Eta model, were compared with  the observational 
data of the CRU and GPCP projects. The comparison shows that the Eta model
reproduces well the main patterns of the summer and winter observed  precipitation
fields  over South America. But the magnitude of precipitation is underestimated
by the model in the  regions of strong convection activity  in summer. This is
probably related to the deficiencies  in the model convection scheme. The larger
underestimation  of observed precipitation by the Eta model driven by HadAM3P than
by the reanalysis  is associated with underestimation of precipitation by HadAM3P.
In winter, the Eta model reproduces better than HadAM3P the magnitude of
precipitation in the equatorial part of the South American continent and position
of ITCZ. The observed number of wet days during  summer season is overestimated by
HadAM3P. The number of wet days in both runs of the Eta model is closer to observations.
During winter, HadAM3P strongly overestimates number of wet days over the continent.
The Eta model driven by reanalysis gives this parameter closer to
observations. The main summer and winter patterns of near surface air temperature 
are reproduced well by both HadAM3P and the Eta model. The Eta model 
overestimates the observed surface  temperature over the central part of the
continent  due to the lack of convective cloudiness in this region. The Eta model
captures observed annual cycle of precipitation in six selected regions over South
America. But the magnitude of precipitation is underestimated in the regions of
strong convection activity in summer. On the whole, these results support the
conclusion that the Eta model with some improvements can be used for downscaling
of the  HadAM3P output fields. The deficiencies of the Eta model in tropics and
subtropics are probably associated with its convection and radiation schemes.

\newpage
\noindent {\bf 1. Introduction}
\bigskip

Climate dynamical downcaling (e.g., Dickinson et al. 1989;  Giorgi and
Bates 1989) is a powerfull method used to get small-scale climatological
features from large-scale climate fields provided by coupled or
atmospheric global models. For the simulation of  current climate the 
regional models are usually integrated for 10-30 years during the period
from 1960 to 2000. The model output fields are evaluated against
observations in order to demonstrate ability  of the models to capture
main features of observed climate.  Good ability of the models in
reproducing regional circulation features and surface variable patterns
does not prove high  accuracy of the model projections of future
climate. But it is necessary step of models validation.  

Multiple regional climate models are used for the climate change
scenario  experiments. For example, the NARCCAP project
(http://www.narccap.ucar.edu)  uses 6 regional climate models for
regional climate projections over  North America. The use of multiple
models for the integration over the same domain allows to evaluate level
of uncertainties  in the simulation of climate, related to different
dynamics and physics of regional models. As it was shown  in (de Elia et
al. 2007), this level  of uncertainties is higher than the level
originated from the model  internal variability or freedom of choice
among configuration parameters. On the opinion of these authors,  an
extremely demanding task is to evaluate uncertainty originated from
variety in the regional climate model physics. 

Till present, only three works about simulations of current 
climate  over South America with regional climate models have been
published.  Seth et al. (2007) simulated South American climate for the
period 1982-2003 using RegCM3 model (Pal et al. 2006) which was  forced
by the boundary conditions derived from the NCAR reanalysis  (Kalnay et
al 1996) and from the integrations of  European-Hamburg (ECHAM) AGCM
(Roeckner et al. 1996). The large-scale meteorological fields of the
regional model driven by the reanalysis boundary conditions and by the
global model boundary conditions were compared with  the driving model
output fields, the reanalysis data, and the Climate  Prediction Center
Merged Analysis of Precipitation (CMAP)  (Xie and Arking 1996) given at
the grid of 2.5 degrees. The results of the comparison indicate that the large
scale climatological features of the regional and global  models are
similar. As compared with observations in some regions  the regional
model can slightly improve the global model performance  or even degrade
its simulation. Solman et al. (2007) performed   regional climate change
experiments over southern South America with the regional MM5 model
(Grell et al. 1993) driven by HadAM3H (Pope et al. 2000) for the period
1980-1990 years. The Climate Research Unit (CRU) dataset  given at
resolution of 0.5 degrees (New et al. 1999) was used for validating monthly
precipitation  and near surface air temperature. The NCEP reanalysis
dataset (Kalney et al. 1996) was used for validating circulation
variables. The authors conclude that the model reproduces the main
regional patterns and seasonal cycle  of surface variables and that 
the model improves the representation of precipitation daily statistics
compared  with the global model.

Pisnichenko and Tarasova (2007) presented one more regional climate 
model developed for regional scale climate simulations and projections
of future climate over South America. The paper demonstrates first
results  of our evaluation of the climate  version (Eta CCS)  developed
from  the Eta regional forecast model (Black 1994). The Eta CCS model
was  integrated over South America for  the period from 1960 to 1990
(current climate) and from 2071 to 2100 (future climate) with the
boundary conditions of HadAM3P  (Pope et al. 2000). The comparison was
made for the current climate between  the large-scale fields of  
geopotencial height, temperature and kinetic energy  at various levels 
of the Eta CCS and HadAM3P outputs.  This comparison demonstrates good
agreement in the field patterns  and absence of trends in time series of
these variables. The spectral distributions of the time series
calculated with Fast Fourier Trasform Algorithm show that the regional
and global models spectras have a high degree of similarity. 

This work shows the results of the Eta model evaluation against  
observations. The shorter period between 1979 and 1985 of current
climate simulation with the Eta CCS model driven by HadAM3P is analyzed.
The results of the Eta CCS model simulation driven by the Reanalysis
boundary conditions are also included in this analysis.  The short 
description of the Eta model, its climate version  Eta CCS, and the 
integration procedure is given in Section 2. Section 3 presents 
observational data sets used for the model evaluation. Section 4 
includes the results of the comparison between the output fields 
of the Eta CCS model, reanalysis, and observational data. Section 4 
gives summary of the results and the conclusions.

\bigskip    

\noindent  {\bf 2. Description of the Eta model and its integration procedure. 
} 

\bigskip

The works of Mesinger et al. (1988), Janjic (1994), and Black (1994) 
present detailed description of the NCEP Eta regional forecast model in
which the planetary boundary layer processes are described by the
Mellor-Yamada level 2.5 model  (Mellor and Yamada 1974). The convective
precipitation scheme is of Betts and Miller (1986) modified by
Janjic(1994). Additional convective scheme  of Kain and Fritsch (1993) was
implemented at the Science Operations Officer/Science and  Training
Resource Center (SOO/STRC)(http://strc.comet.ucar). The shortwave  and
longwave radiation  schemes use the parameterizations of Lacis and Hansen
(1974) and  Fels and Schwartzkopf (1975),  respectively. The land-surface
scheme is of Chen et al. (1997).  The grid-scale  cloud cover fraction is 
parameterized as a function of relative humidity and cloud  water (ice)
mixing ratio (Xu and Randall 1996; Hong et al. 1998). Convective  cloud
cover fraction is parameterized as a function of precipitation rate (Slingo
1987).

We made multiple changes and corrections in the workstation (WS) Eta 
modeling package (version of 2003) developed at 
SOO/STRC(http://strc.comet.ucar) during the preparation of the climate 
version Eta CCS. These modifications  include restart program, SST and SICE 
data updating day integrated output etc. The full description of these 
modifications is given in (Pisnichenko and Tarasova, 2007). The new shortwave 
radiation scheme(CLIRAD-SW-M) of Chou and Suarez (1999) and 
Tarasova and Fomin (2000) and longwave radiation  scheme of Chou et al.(2001) 
were implemented in the Eta CCS model as options. 

In the first experiment (Had - Eta CCS), the Eta CCS model was forced   at
its lateral and bottom boundary by the output of HadAM3P atmospheric 
global model. The HadAM3P used  SST, SICE (sea ice) and greenhouse  gases 
and aerosol concentration as external driving from the coupling global
model HadCM3.  In the second experiment (R2 - Eta CCS), the Eta CCS model was
driven by the boundary conditions derived fr om the NCEP-DOE AMIP-II
reanalysis (R2) dataset (Kanamitsu et al. 2002). The Eta CCS model received
lateral boundary conditions every 6 hours  and SST and SICE  data every  15
days and then interpolated them at each  time step. The climate mean values
of soil moisture and soil temperature  were used as initial conditions. The
first year of the integration was  considered as spin-up period related  to
the stabilization of soil  variables and was not used in the analysis. 

The area of integration was centered at  $58.5^{\circ}$ W longitude and 
$22.0^{\circ}$ S latitude  and covers the territory of South American
continent with adjacent oceans ($55^{\circ}$ S - $16^{\circ}$ N,
$89^{\circ}$ W - $29^{\circ}$ W).  The model was integrated on the
211$\times$115 horizontal grid with grid spacing of 37 km.  The 38 eta
vertical coordinate layers were  used. For the modern climate integration
the Betts-Miller cumulus convection parametrization scheme and the ETA
model original shortwave and longwave radiation schemes were chosen.

\bigskip    

\noindent  {\bf 3. The observational data used for the Eta model validation
} 

\bigskip

The data of Climatic Research Unit (CRU) of the University of East
Anglia is used for the evaluation of precipitation and surface
temperature simulated by both regional and global models.  The CRU
datasets were created from the station data interpolated as a function 
of latitude, longitude and elevation. In this work, the CRU CL 2.0
monthly mean dataset of Mitchell et al. (2003) given at the  horizontal
resolution  of 0.5 degrees for the period from 1901 to 2000 is used.  

The R2 reanalysis data set (Kanamitsu et al. 2002) is used  for the
comparison with the model-simulated fields of temperature,  geopotential
height, and kinetic energy at the model levels. This dataset is  given at
the resolution of 2.5 degrees for the period from  1979  through the
previous year. This quasi-observational dataset is based on the 
observational data obtained from various data sources, checked, and then 
assimilated  by a data assimilation system. In the data-sparse areas (for
example, in the tropics) the reanalysis data set is more model-dependent.   

The GPCP precipitation data set based  on the  combination of satellite
estimates and gauge observations  are used for the evaluation   of the Eta
CCS and HadAMP3 model-generated precipitation. Its temporal  coverage is
from 1979 through present, while the spatial coverage is  2.5-degree global
grid. The monthly mean (Version 2) and 5-day   intervals (Pentad) products
(Adler et al. 2003; Xie et al. 2003)  are selected for the model
validation.

\bigskip    

\noindent  {\bf 4. The Eta model evaluation against observations
} 

\bigskip 

\noindent {\it a. Summer (DJF) and winter (JJA) mean precipitation and temperature 
patterns simulated by (Had - Eta CCS) and (R2 - Eta CCS) as compared with the CRU 
and GPCP data sets  }
\bigskip

Fig. 1 compares mean precipitation in CRU observations, GPCP data sets, HadAM3P, 
and (R2 - Eta CCS), (Had - Eta CCS) model simulations for austral summer months  from
December,  to February (DJF) averaged over 4 years from 1980 to 1983. These
months are characterized  by strong precipitation in most South America east of
the Andes. Both CRU and GPCP data sets  reproduce precipitation maximum of 8
mm/day associated with the South Atlantic  Convergence Zone (SACZ), while both
the HadAM3P and (Had - Eta CCS) model runnings are not able  to reproduce this maximum. The
(R2 - Eta CCS) model running, which is driven by quasi-observational  boundary conditions of
Reanalysis II, captures this maximum but over smaller region.  The precipitation
amount area of 4 mm/day seen in the observational data sets  is well captured by
HadAM3P, while this area is underestimated in the Eta CCS  simulations. The model
failure in reproducing precipitation patterns associated  with SACZ can be
related to the deficiencies in their radiation and convection schemes  (Tarasova
et al., 2006; Figueroa et al., 2006) or/and to the deficiencies  in the model
humid transport. All models overestimate precipitation amount over high
latitudes. As it was discussed by Solman et al. (2007) this can be related to the
deficiencies in regional climate modeling of convection processes over mountains. 
Collection of surface precipitation data  over elevated terrains also is not perfect.

Figure 2 shows the same comparison of the precipitation fields for the austral
winter  months that is June, July and August (JJA) in South America. Both CRU and
GPCP observational  data sets demonstrate precipitation maximum over northern
part of the continent but its  amount is larger in the CRU data set. The largest
difference between the model results is also seen over this area where the
regional model demonstrates better representation of the  precipitation amount
than the global model. (Had - Eta CCS) is also better captures  the location of the
Inter Tropical Convergence Zone (ITCZ) than HadAM3P as compared with  the GPCP
data set. The dry conditions over the South American continent are reproduced 
well by all models. All models overestimate observational precipitation amount
over       southern Chile. The observed bias in the model-simulated  mean
precipitation fields may be caused by the deficiencis in the Betts-Miller 
convection scheme used in these simulations or by the deficiencies in the model 
humid transport.

The difference in precipitation amount between the models and observational
data  sets is related to the difference in precipitation frequency or in
precipitation  event intencity. The precipitation frequency is 
defined as  number of wet days per month or per season. The wet day is
defined in the  CRU data set as a day with a  daily precipitation amount
larger than 0.1 mm. Fig. 3 presents number of wet days during summer (DJF),
averaged over the period from 1980 to 1983.  The CRU observations and the
HadAM3P,  (R2 - Eta CCS), (Had - Eta CCS) simulation results are compared. The
comparison shows that  during summer the model runnings overestimate number of wet
days over the South American  continent as compared with the CRU
observations. The largest overestimation of the area  of 80 wet days per
season is seen over north part of the continent in the HadAM3P simulations.
The (R2 - Eta CCS) model running simulates better the area of 80 days per season than 
HadAM3M and (Had - Eta CCS). All model runnings overestimate the number of the wet days
in the  central part of the continent by 20 wet days per season. Fig. 4 shows
the comparison  of the number of wet days during austral winter (JJA). All
models overestimate rainy  day frequancy in the northern part of the
continent by 20 wet days per season and large overestimation is seen over the
Andes. 

Figs. 5 and 6 present the fields of the  near surface air temperature at 2 m in
CRU observations, GPCP data sets, HadAM3P,  (R2 - Eta CCS), (Had - Eta CCS) model
simulations averaged for austral summer (DJF) and winter (JJA) months and 4 years
from 1980 to 1983. All models reproduce main patterns of the temperature  field
over South America but the regional model overestimates temperature magnitude 
over the northern and central parts of the continent by 2-3 C in summer. This is
probably  associated with the lack of convective  precipitation and hence
convective cloudiness in these regions in the Eta model that related to the
deficiencies in the model  convection parameterization scheme. Temperature values
smaller  by 0.5-1.0 C were obtained over SACZ in the Eta model simulations 
with new  solar  radiation scheme for January 2003 (Tarasova et al., 2006) as
compared with its original scheme.   For the winter months (JJA) the HadAM3P
model shows temperature maximum of 27 C in  the  north-east part of the continent
which is not seen in the CRU data set. The same   maximum is captured by the
(Had - Eta CCS) model running that is probably related to the impact of the global model
boundary conditions.

\bigskip
\noindent {\it b. Annual cycle of precipitation for the selected regions and its 
interannual variations simulated by (Had - Eta CCS), (R2 - Eta CCS) and HadAM3P as 
compared  with the GPCP and CRU data }
\bigskip

We examined area averaged precipitation annual cycle over the following
regions: Amazon (AM), North Amazon (NA), North East Brazil (NE), South Brazil
(SB),  South East Brazil (SE), Pantanal (PA)  shown in Fig. 7. These regions
have distinctive climate features  that differ from one region to another. 
For these regions we calculated area averaged monthly mean daily
precipitation (mm\,d$^{-1}$) averaged over 1980-1983 years for  
the CRU and GPCP observational dat sets as well as the (R2 - Eta CCS), HadAM3P,
(Had - Eta CCS) simulation results. Only land grid points  were taken into
account in the calculations. Both observational data sets  demonstrate
similar values of monthly mean precipitation for all regions.  According to
observational estimates, in the Amazon region the largest values of 
precipitation are from January to Abril and the smallest values are from 
June to October. All models capture observed annual cycle of precipitation
while the magnitude of precipitation is  underestimated, particularly in the
summer months.  In the North Amazon region there is a distinctive peak of
precipitation in Abril which is captured by both HadAM3P and (Had - Eta CCS)
models but its magnitude is  underestimated. The (R2 - Eta CCS)
model captures precipitation maximum in June. In the North East region a peak
of precipitation in observational data sets is seen from February to March.
All models underestimate precipitation during these months. The monthly mean
precipitation in the South Brazil region has 3 maximums, in February, March
and November. The shape of annual cycle of precipitation is well captured by
the (R2 - Eta CCS) model. Both HadAM3P and (Had - Eta CCS) models capture the
precipitation maximum in February but underestimate precipitation values
during all months. A strong annual cycle is observed in the South East and
Pantanal regions and all models  reproduce its shape. In the PA region the
magnitude of precipitation is  underestimated by all models in the summer
months.

Figs. 9 and 10 show precipitation anomaly averaged over the same regions (Fig.
7)  for summer (DJF) and winter (JJA) seasons, respectively. The precipitation
anomalies  for each year from 1980 to 1983 were calculated as a difference
between the season mean precipitation for each year and that averaged over 4
years. In the Pantanal region in summer all models have most 
difficulties in simulating  interannual variation of season mean precipitation. 
Note, that both observational  data sets also have largest differences  
over this region.  The best agreement between the 
model-simulated and observed precipitation anomalies is seen over the AM and 
SB regions in summer and over the AM, NE, SB and PA regions in winter.

\bigskip    

\noindent  {\bf 5. Conclusions 
} 

\bigskip

The output fields of precipitation, precipitation frequency, and near surface air
temperature simulated by the Eta model over South America  were compared  with the
observational data of the CRU and GPCP projects. The runs of the Eta model  were
made for the period from January 1979 to December 1983 using boundary conditions 
derived from the reanalysis and outputs of the HadAM3P model.  The model  monthly
mean results were averaged over four years from January 1980  to December 1983. The
comparison of the model-simulated and observed precipitation fields shows that  the
main  patterns of observed fields are reproduced reasonably well by the Eta model
driven by both sets of boundary conditions.  In both Eta model experiments, the
magnitude of precipitation in summer is underestimated in the  regions of strong
convection activity  related to South Atlantic Convection Zone. It can be caused by
the deficiencies in the model convection and radiation  schemes and in the model
humid transport. The magnitude of the underestimation is larger in the experiments
used the Eta model driven by HadAM3P than the Eta model  driven by reanalysis. This
is associated with the lack of precipitation in the HadAM3P model in this region in
summer. In winter, the dry conditions over South America  are reproduced well by
both HadAM3P and the Eta model. The Eta model driven by HadAM3P reproduces better 
than HadAM3P the magnitude of precipitation in the equatorial part of the South
American continent and the  position of ITCZ.   

The comparison of the model-simulated and observed precipitation frequencies  shows
that during summer season the observed number of wet days (CRU)  is overestimated by
HadAM3P. The overestimation is larger over the  northern part of the continent. The
precipitation frequency is closer to observations in both runs of the Eta model,
particularly in the run with  the boundary conditions derived from the reanalysis. 
During winter season, HadAM3P strongly overestimates the precipitation frequency
over all continent. The Eta model driven by reanalysis simulates the precipitation
frequency closer to observations.  The observed patterns of near surface air
temperature (CRU)  are reproduced well by the HadAM3P and Eta models in both summer 
and winter seasons. In summer, the Eta model overestimates the magnitude of  surface
temperature over the central part of the continent due to the lack of convective
cloudiness in this region. In winter, both HadAM3P and the Eta model driven by
HadAM3P shows the temperature maximum in the northern-eastern part  of the continent
which is not present in the CRU data. The Eta model captures the observed annual
cycle of precipitation in the six selected regions over South America. But the
magnitude of precipitation is underestimated in the regions of strong convection
activity in summer. On the whole, these results support the conclusion that the Eta
model with some improvements can be used for downscaling of the  HadAM3P output
fields. The deficiencies of the Eta model in tropics and subtropics are probably
associated with the deficiencies in the model convection and radiation schemes.

\newpage
\centerline{{\bf References }}
\bigskip

\setlength{\parindent}{-1.0\parindent}

Adler RF, Huffman GJ, Chang A, Ferraro R, Xie P, Janowiak J, Rudolf B, 
Schneider U, Curtis S, Bolvin D, Gruber A, Susskind J, Arkin P (2003) 
The Version 2 Global Precipitation Climatology Project (GPCP) Monthly 
Precipitation Analysis (1979-Present). J Hydrometeor 4:1147-1167

Betts AK, and Miller MT (1986) A new convective adjustment 
scheme. Part II: Single column tests GATE wave, BOMEX, and 
Arctic air-mass data. Quart J Roy Met Soc 112: 693-703

Black TL (1994) NMC notes: the new NMC mesoscale Eta model: 
description and forecast examples. Wea Forecasting 9:256-278

Chen FK, Janjic Z, and Mitchel K, (1997) Impact of the 
atmospheric surface-layer parameterizations in the new 
land-surface scheme of the NCEP mesoscale Eta model. Bound-Layer Meteor 
85: 391-421  
 
Chou M-D, and Suarez MJ  (1999) A solar radiation 
parameterization (CLIRAD-SW) for atmospheric studies. Preprint  
NASA/Goddard Space Flight Center, Greenbelt, Maryland, 38 pp

Chou M-D, Suarez MJ, Liang X-Z, and Yan M M-H (2001) A thermal 
infrared radiation parameterization for atmospheric Studies.
Preprint NASA/Goddard Space Flight Center, Greenbelt, Maryland, 55 pp

Dickinson RE, Errico RM, Giorgi F, and Bates GT (1989)  
A regional climate model for the western United States. 
Clim Change 15:383-422

de Elia R, Caya D, Cote H, Frigon A, Biner S, Giguere M, 
Paquin D, Harvey R, Plummer D (2007) Evaluation of uncertainties 
in the CRCM-simulated North American climate. Clim Dyn, in Press 

Fels SB, and Schwartzkopf MD (1975) The simplified exchange 
approximation: A new method for radiative transfer 
calculations. J Atmos Sci 32:1475-1466

Giorgi F, and Bates GT (1989) The climatological skill of a regional 
model over complex terrain. Mon Wea Rev 117:2325-2347

Grell GA, Dudhia J, Stauffer DR (1993) A description of the 
fifth-generation Penn System/NCAR mesoscale model (MM5). 
NCAR Tech Note NCAR/TN-398+1A, 107 pp 

Hong S-Y, Yuang H-M, and Zhao Q (1998) Implementing of 
prognostic cloud  scheme for a regional spectral model. Mon Wea Rev 
126:2621-2639

Janjic ZI (1994) The step-mountain eta coordinate model: further 
development of the convection, viscous sublayer, and 
turbulence closure schemes. Mon Wea Rev 122:927-945

Kain JS, and Fritsch JM (1993) A one-dimensional entraining detraining 
plume model and its applications in convective parameterization.
J Atmos Sci 23:2784-2802 

Kanamitsu M, Ebisuzaki W, Woollen J, Yang S-K, Hnilo JJ, 
Fiorino M, and Potter GL (2002) NCEP-DOE AMIP-II Reanalysis 
(R-2). Bull Amer Meteor Soc 83:1631-1643

Kalnay E, Coauthors (1996) The NCEP-NCAR 40-year reanalysis 
project. Bull Am Meteor Soc 77:437-471

Lacis AA, and Hansen JE (1974) A parameterization for the 
absorption of solar radiation in the Earth's atmosphere. J Atmos Sci 
31:118-133

Mellor GL, and Yamada T (1974) A hierarchy of turbulence 
closure models for boundary layers. J Atmos Sci 31:1791-1806.

Mesinger F, Janjic ZI, Nickovic S, Gavrilov D, and Deaven DG (1988) The
step-mountain coordinate: model description and performance for cases of
Alpine lee cyclogenesis and for a case of Appalachian redevelopment. Mon
Wea Rev 116:1493-1518 

Mitchell TD, Carter TR, Jones PD, Hulme M, New M (2003) A comprehensive 
set of high-resolution grids of monthly
climate for Europe and the globe: the observed record (1901-2000) and 16 scenarios
(2001-2100). J Clim 

New MG, Hulm M, Jones PD (1999) Representing twentieth-century space time
climate variability. Part I. Development of a 1961-1990 mean monthly 
terrestrial climatology. J Clim 12:829-856

Pal JS, Giorgi F, Bi X et al (2006) The ICTP RegCM3 and RegCNET: 
regional climate modeling for the developing World. 
Bull Am Meteorol Soc (in press)

Pope V, Gallani M, Rowntree P, Stratton R (2000) The impact of 
new physical parameterizations in the Hadley Centre Climate model.
Clim Dyn 16:123-146

Pisnichenko IA, and Tarasova TA (2007)  The climate version of the 
Eta regional forecast model. 1. Evaluation of consistency 
between the Eta model and HadAM3P global model.
http://arxiv.org/abs/0709.2110v2  
	
Roeckner E, Aroe K, Bengtsson L et al (1996) The atmospheric general 
circulation model ECHAM-4: model description and simulation of 
present day climate. Technical report, 218, Max-Plank Institute for 
Meteorology

Seth A, Rauscher SA, Camargo SJ (2007) RegCM3 regional climatologies
for South America using reanalysis and ECHAM global model driving fields.
Clim Dyn 28:461-480 doi: 10.1007/s00382-006-0191-z

Slingo JM (1987) The development of a cloud prediction model 
for the ECMWF model. Quart J Royal Met Soc 113:899-927 

Solman SA, Nunez MN, Cabre MF (2007) Regional climate change experiments 
over southern South America. I: present climate. Clim Dyn, in Press

Tarasova TA, and Fomin BA (2000) Solar radiation absorption 
due to water vapor: Advanced broadband parameterizations. J Appl Meteor 
39:1947-1951

Xie P, Janowiak J, Arkin PA, Adler RF, Gruber A, Ferraro RR, Huffman GJ,
Curtis S (2003) GPCP Pentad Precipitation Analysis: Am Experimental 
Dataset Based on Gauge Observations and Satellite Estimates. J Clim, 
16:2197-2214

Xie P, Arkin P (1996) Analesis of global monthly precipitation using 
gauge observation, satellite estimates and numerical model prediction. 
J Clim 9:840-858

Xu K-M, and Randall DA (1996) A semi empirical cloudiness
parameterization  for use in climate models. J Atmos Sci 
53:3084-3102

\newpage 

\centerline{{\bf Figure captions}} 

\bigskip

\noindent {\bf Figure~1.} DJF mean precipitation (mm\,d$^{-1}$) averaged 
over 1980-1983 years: (a) GPCP, (b) CRU, (c) HadAM3P, (d) 
Had-Eta CCS, (e) R2-Eta CCS. 

\noindent {\bf Figure~2.} JJA mean precipitation (mm\,d$^{-1}$) averaged 
over 1980-1983 years: (a) GPCP, (b) CRU, (c) HadAM3P, (d) 
Had-Eta CCS, (e) R2-Eta CCS. 

\noindent {\bf Figure~3.} Number of wet days during austral summer (DJF) averaged 
over 1980-1983 years: (a) CRU, (b) HadAM3P, (c) Had-Eta CCS, (d) R2-Eta CCS.

\noindent {\bf Figure~4.} Number of wet days during austral winter (JJA) averaged 
over 1980-1983 years: (a) CRU, (b) HadAM3P, (c) Had-Eta CCS, (d) R2-Eta CCS.

\noindent {\bf Figure~5.} DJF mean near surface air temperature ($^{\circ}$C) averaged 
over 1980-1983 years: (a) CRU, (b) HadAM3P, (c) Had-Eta CCS.

\noindent {\bf Figure~6.} JJA mean near surface air temperature ($^{\circ}$C) averaged 
over 1980-1983 years: (a) CRU, (b) HadAM3P, (c) Had-Eta CCS.

\noindent {\bf Figure~7.} The 6 regions selected for the analysis: 
Amazon (AM),North Amazon (NA), North East Brazil (NE), South Brazil (SB), 
South East Brazil (SE), Pantanal (PA).

\noindent {\bf Figure~8.} Annual cycle of monthly mean daily precipitation (mm\,d$^{-1}$)
averaged over 1980-1983 years and 6 regions (Amazon (AM), North Amazon
(NA), North East Brazil (NE), South Brazil (SB), South East Brazil (SE), 
Pantanal (PA)) from CRU (dashed), GPCP (dot-dashed), R2-Eta CCS (square),  
HadAM3P (filled circle), Had-Eta CCS (triangle).

\noindent {\bf Figure~9.} DJF mean precipitation anomaly (mm\,d$^{-1}$)
averaged over 1980-1983 years and 6 regions (Amazon (AM), North Amazon
(NA), North East Brazil(NE), South Brazil (SB), South East Brazil(SE), 
Pantanal (PA)) from CRU (dashed), GPCP (dot-dashed), R2-Eta CCS (square), 
HadAM3P (filled circle), Had-Eta CCS (triangle).

\noindent {\bf Figure~10.} JJA mean precipitation anomaly (mm\,d$^{-1}$)
averaged over 1980-1983 years and 6 regions (Amazon (AM), North Amazon
(NA), North East Brazil (NE), South Brazil (SB), South East Brazil(SE), 
Pantanal (PA)) from CRU (dashed), GPCP (dot-dashed), R2-Eta CCS (square), 
HadAM3P (filled circle), Had-Eta CCS (triangle).    
\newpage
\begin{figure}[p]
\centerline{
      \includegraphics[angle=270, width=53mm]{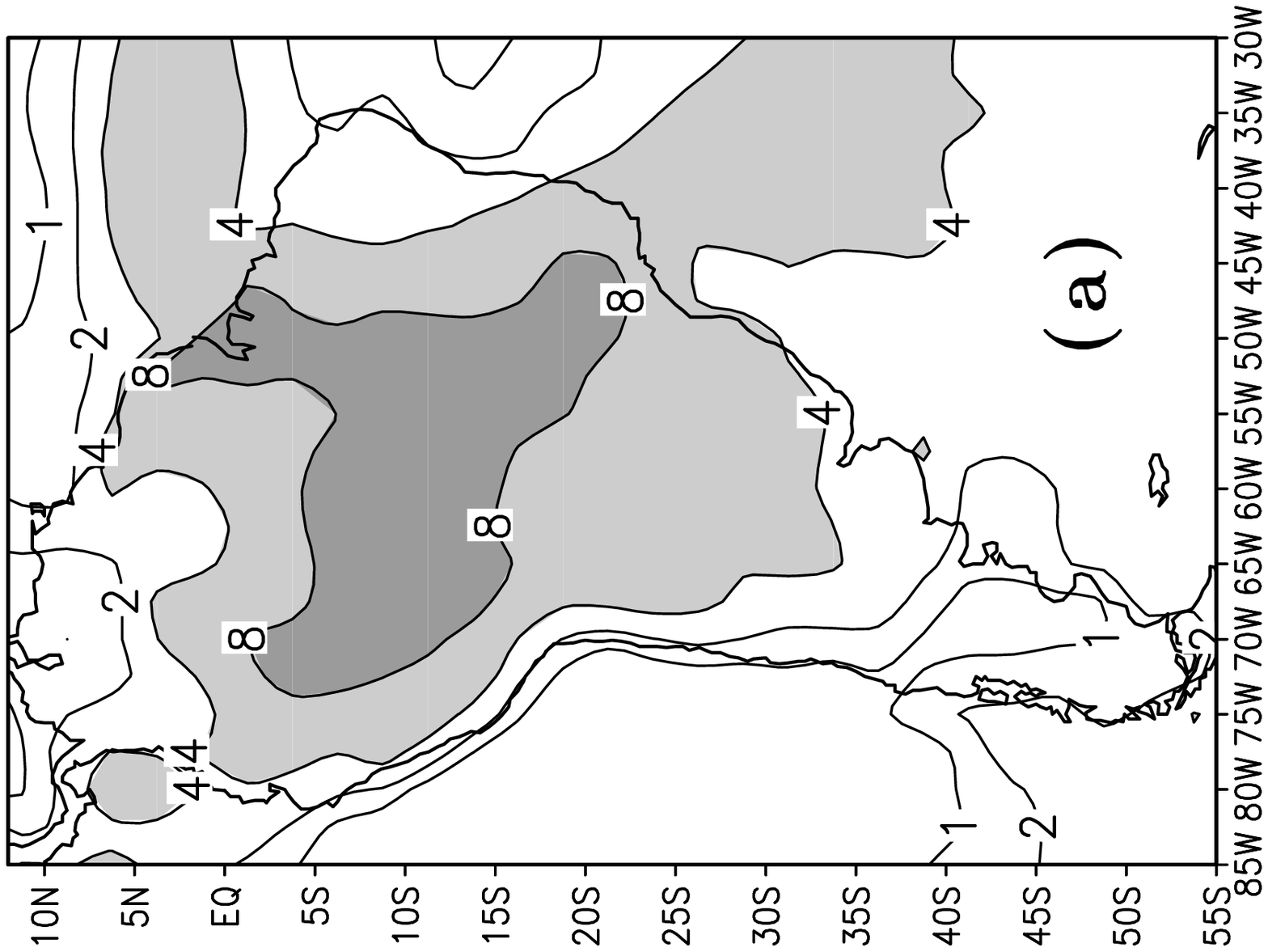}      
      \includegraphics[angle=270, width=53mm]{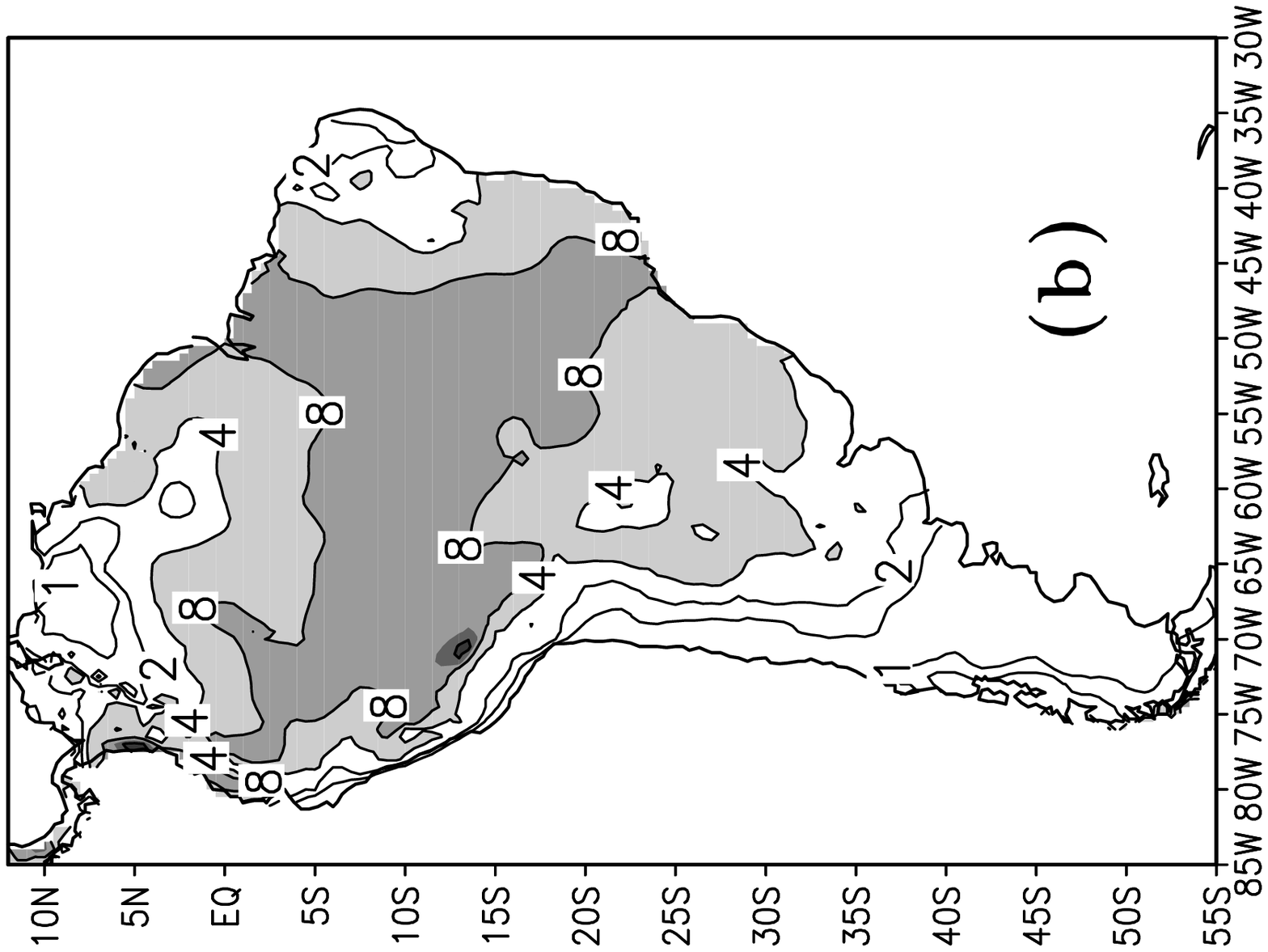}
      \includegraphics[angle=270, width=53mm]{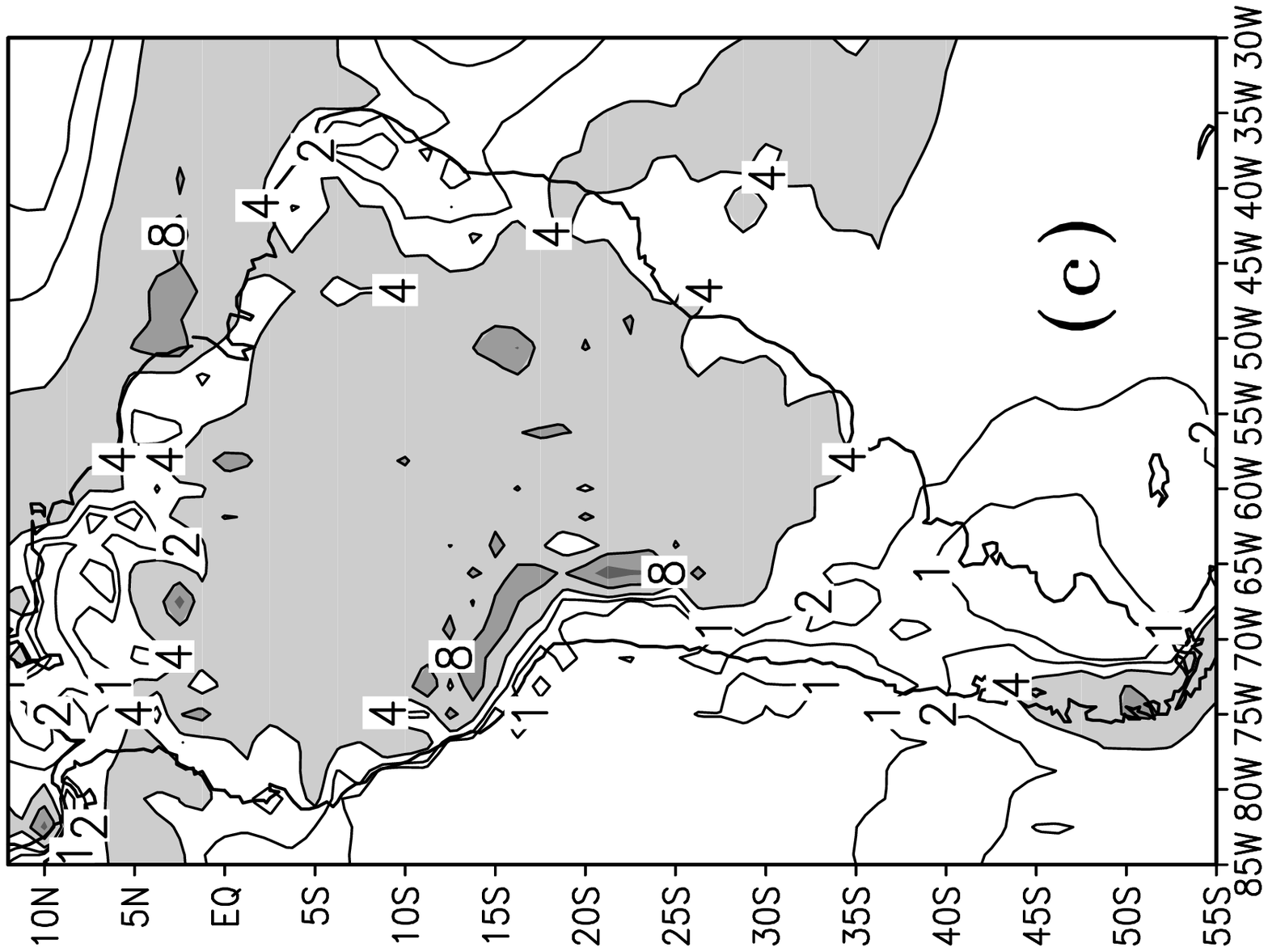}      
      }

\centerline{
      \includegraphics[angle=270, width=53mm]{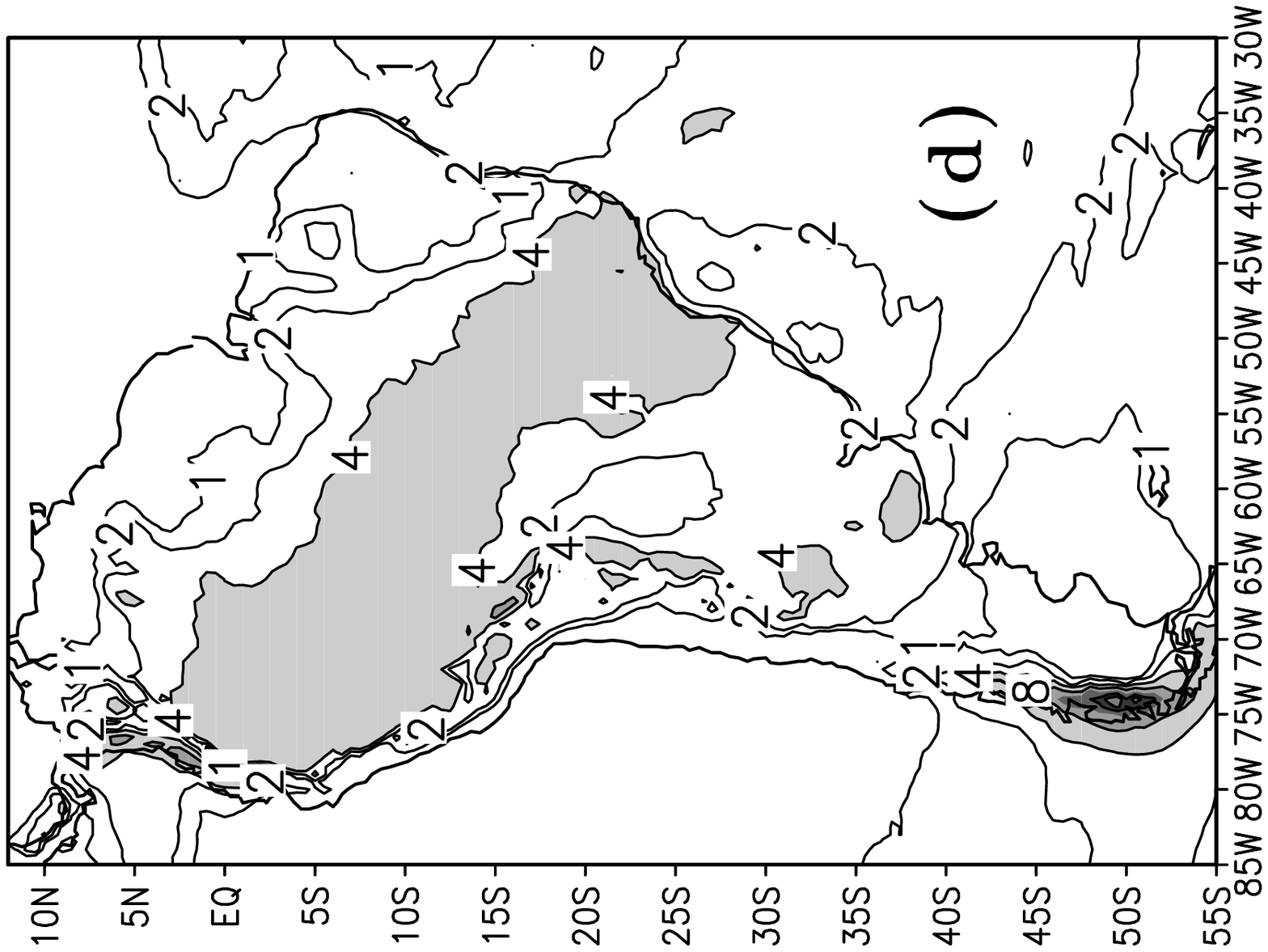}
      \includegraphics[angle=270, width=53mm]{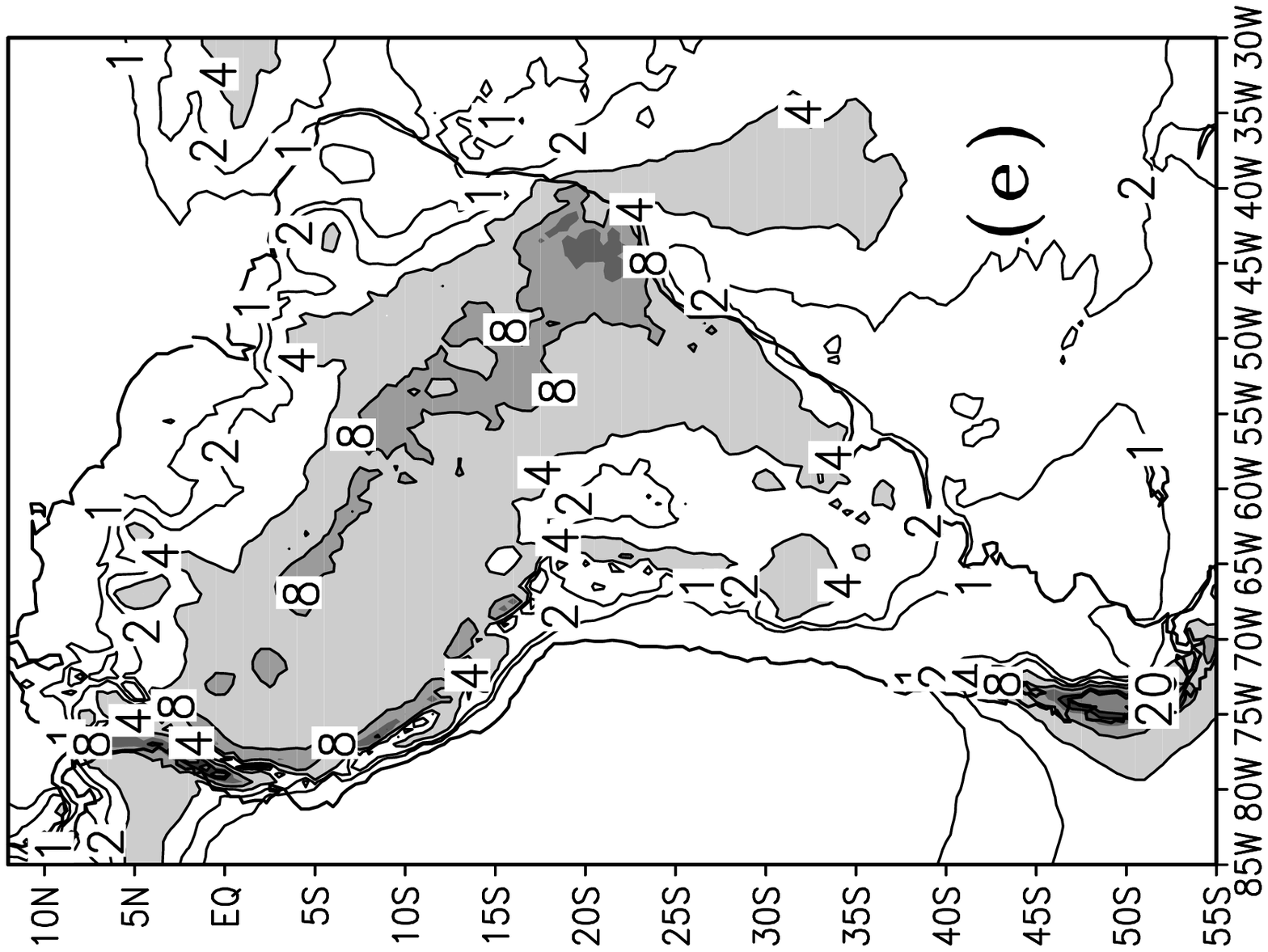}      
      }
 
\bigskip     
\caption[1]{\label{1} DJF mean precipitation (mm\,d$^{-1}$) averaged 
over 1980-1983 years: (a) GPCP, (b) CRU, (c) HadAM3P, (d) 
Had-Eta CCS, (e) R2-Eta CCS. 
      }  
\end{figure}

\begin{figure}[p]
\centerline{
      \includegraphics[angle=270, width=53mm]{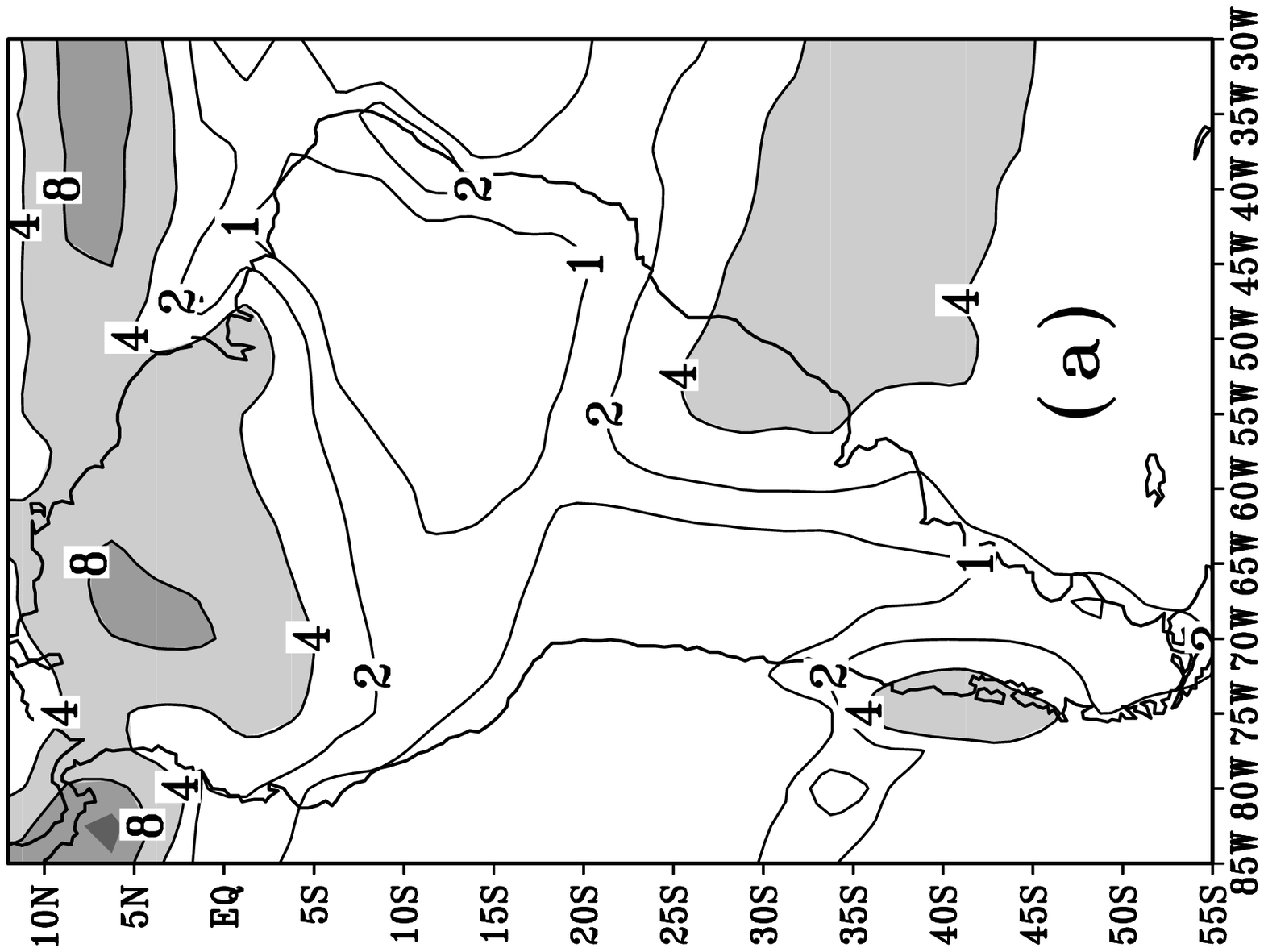}      
      \includegraphics[angle=270, width=53mm]{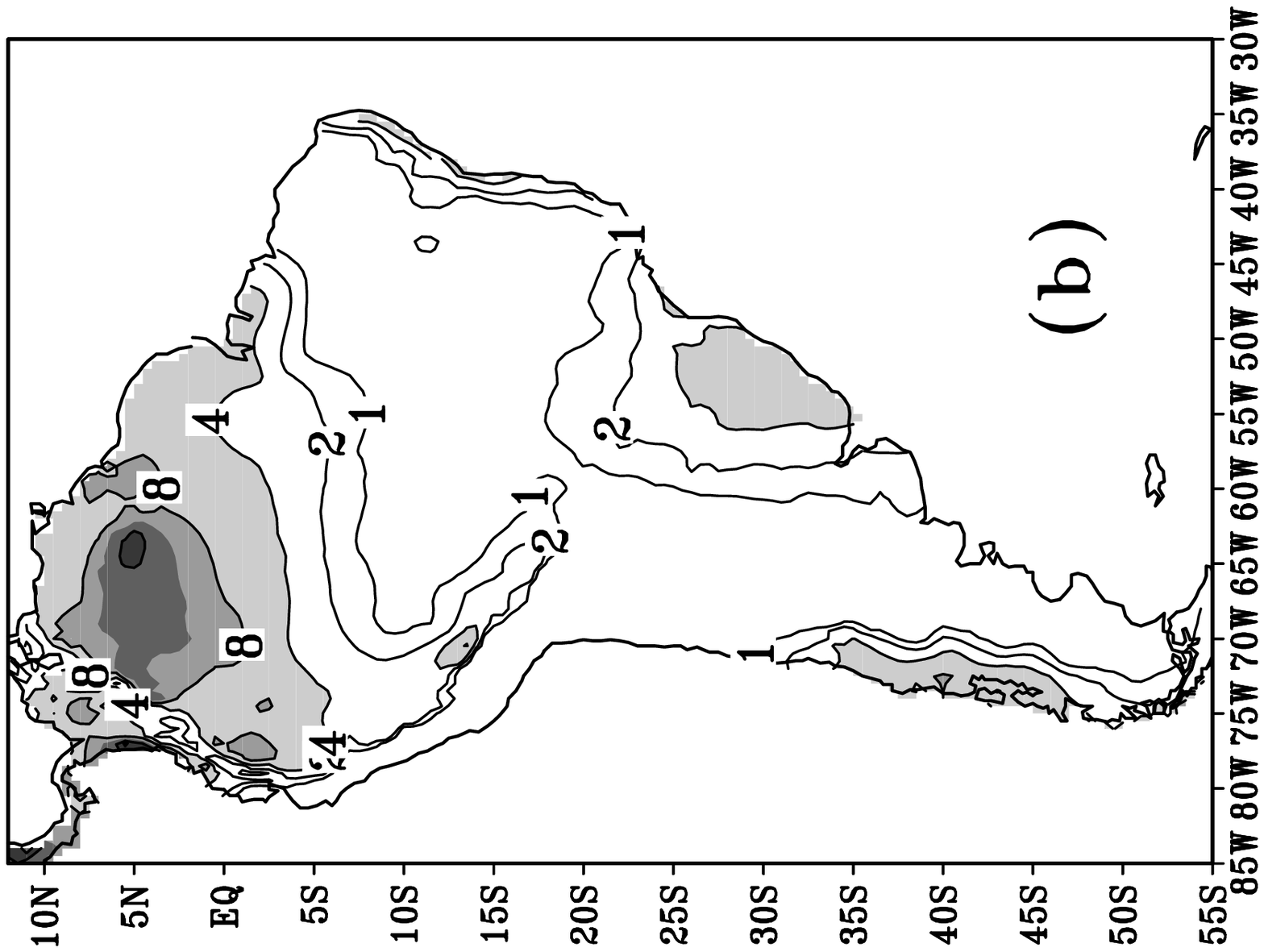}
      \includegraphics[angle=270, width=53mm]{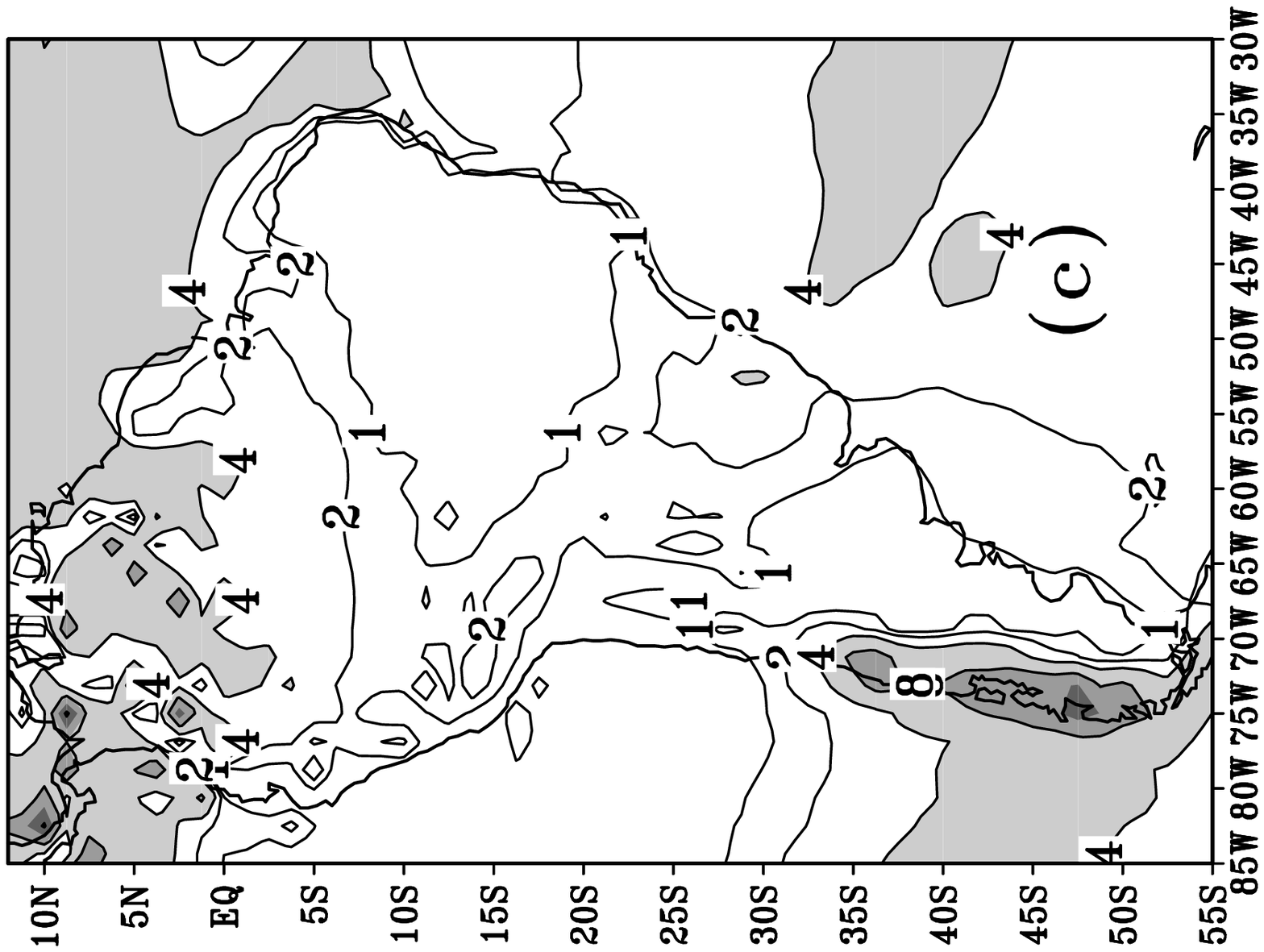}      
      }

\centerline{
      \includegraphics[angle=270, width=53mm]{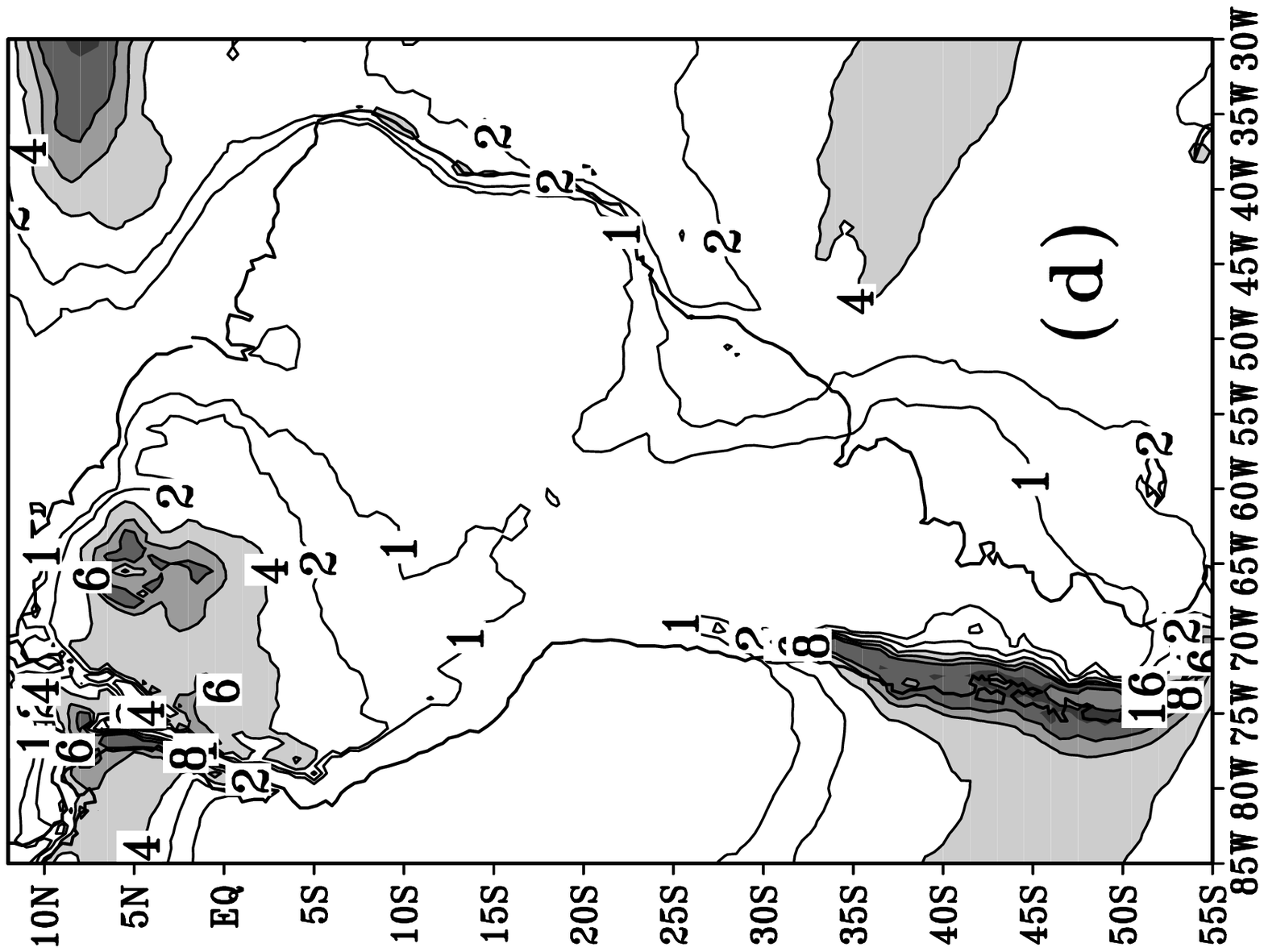}
      \includegraphics[angle=270, width=53mm]{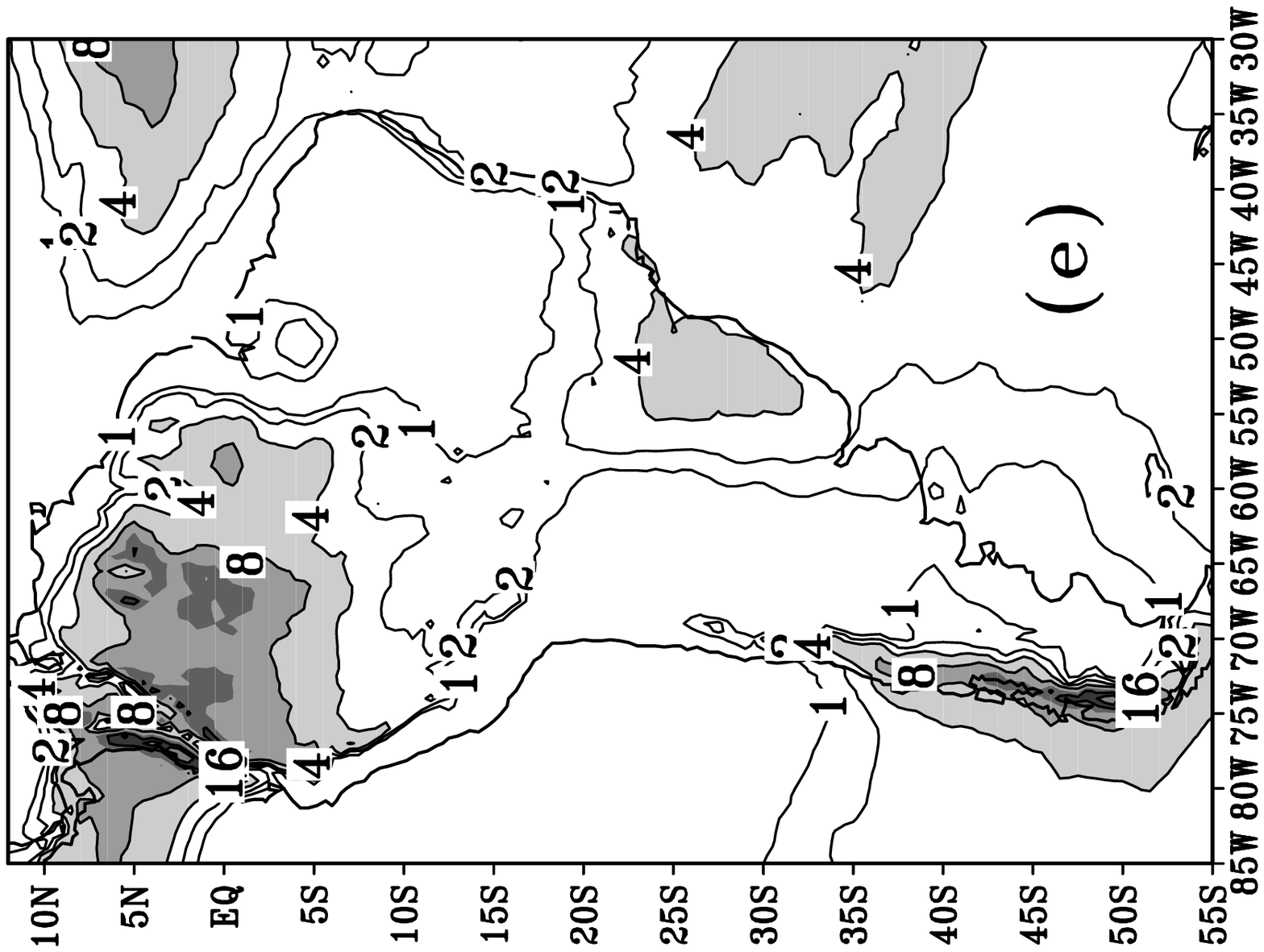}      
      }
 
\bigskip     
\caption[1]{\label{1} JJA mean precipitation (mm\,d$^{-1}$) averaged 
over 1980-1983 years: (a) GPCP, (b) CRU, (c) HadAM3P, (d) 
Had-Eta CCS, (e) R2-Eta CCS. 
      }  
\end{figure}
\begin{figure}[p]
\centerline{
      \includegraphics[angle=270, width=60mm]{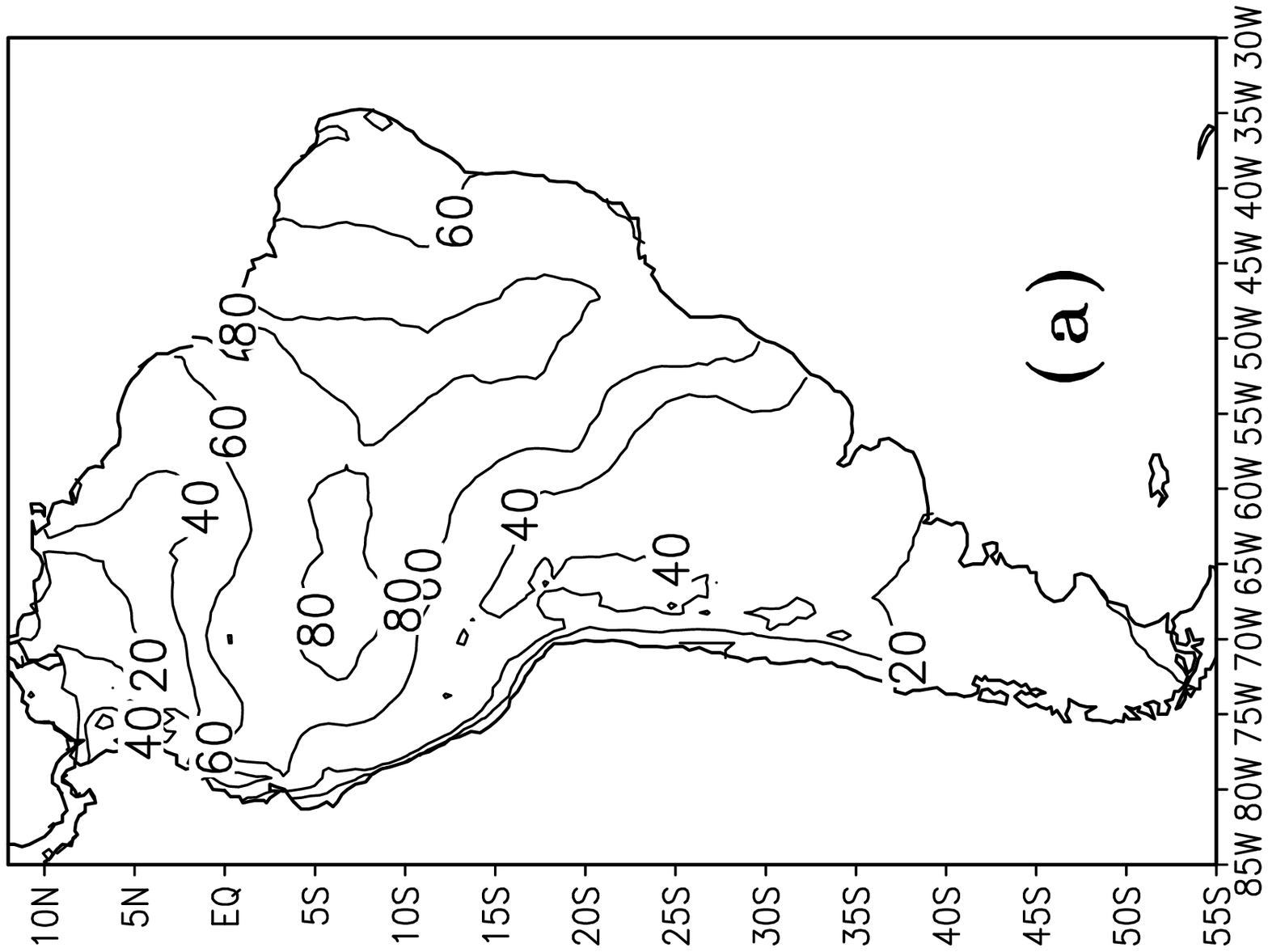}      
      \includegraphics[angle=270, width=60mm]{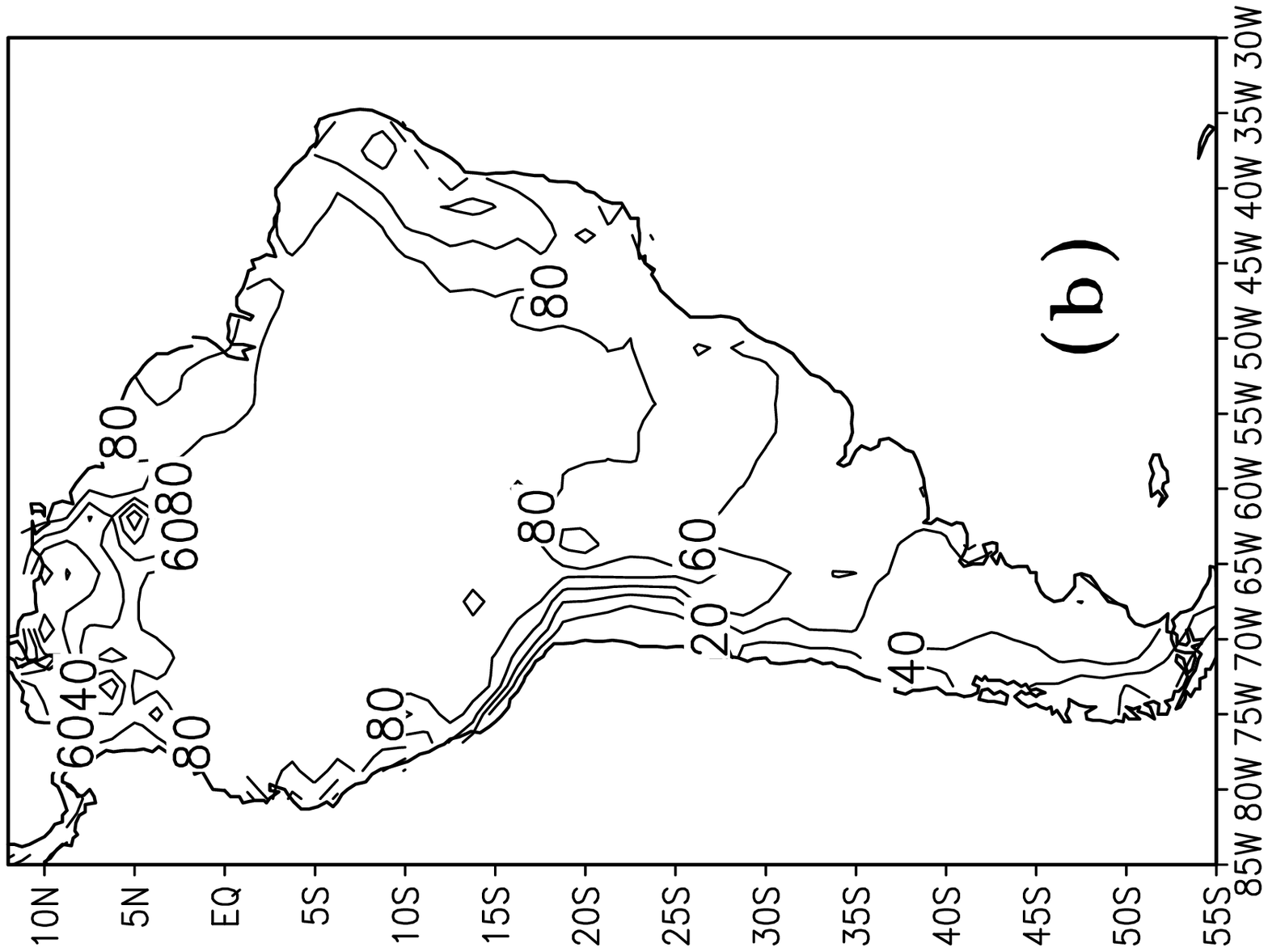}      
      }

\centerline{
      \includegraphics[angle=270, width=60mm]{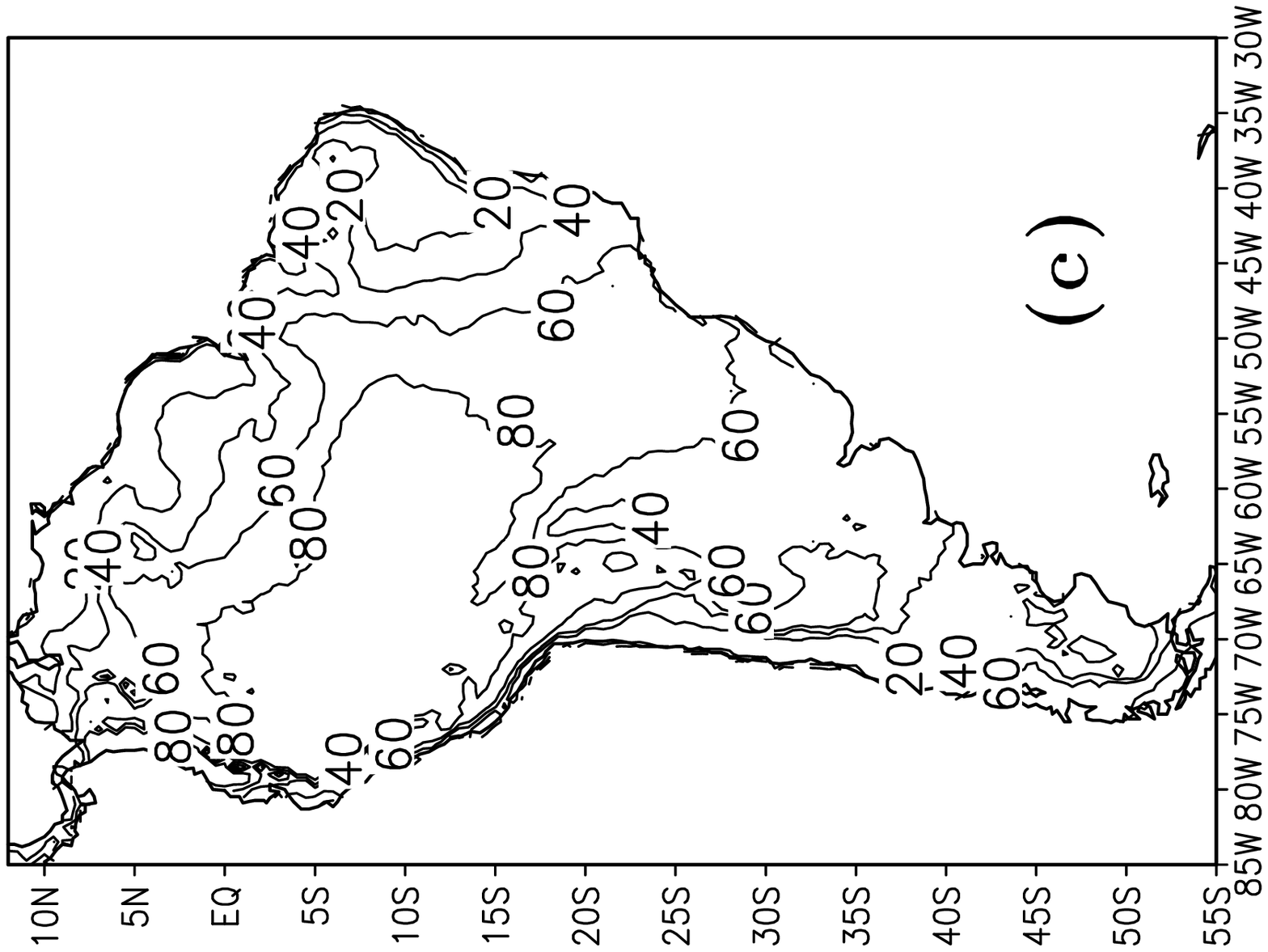}
      \includegraphics[angle=270, width=60mm]{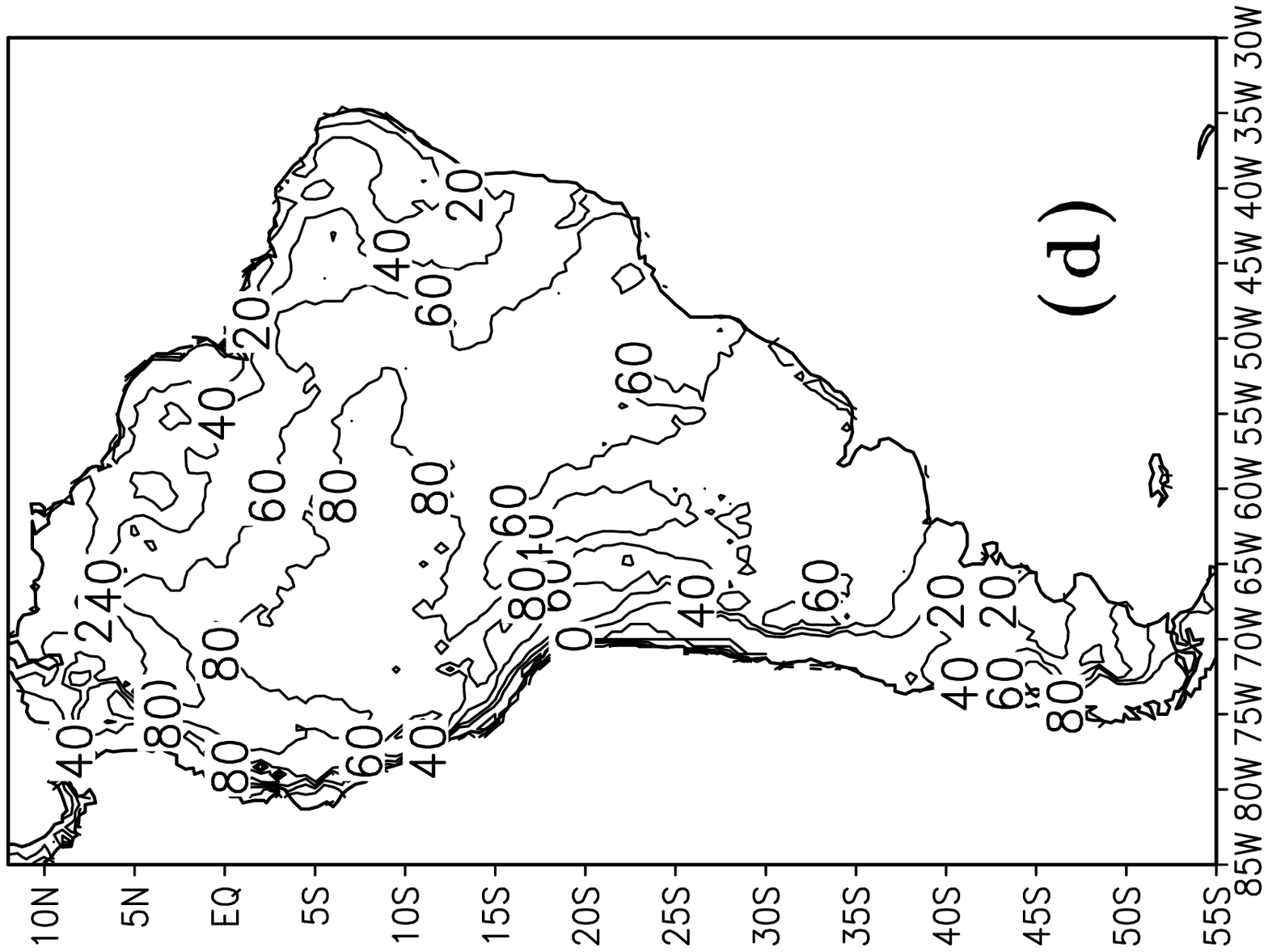}      
      }
 
\bigskip     
\caption[1]{\label{1} Number of wet days during austral summer (DJF) averaged 
over 1980-1983 years: (a) CRU, (b) HadAM3P, (c) 
Had-Eta CCS, (d) R2-Eta CCS. 
      }  
\end{figure}
\begin{figure}[p]
\centerline{
      \includegraphics[angle=270, width=60mm]{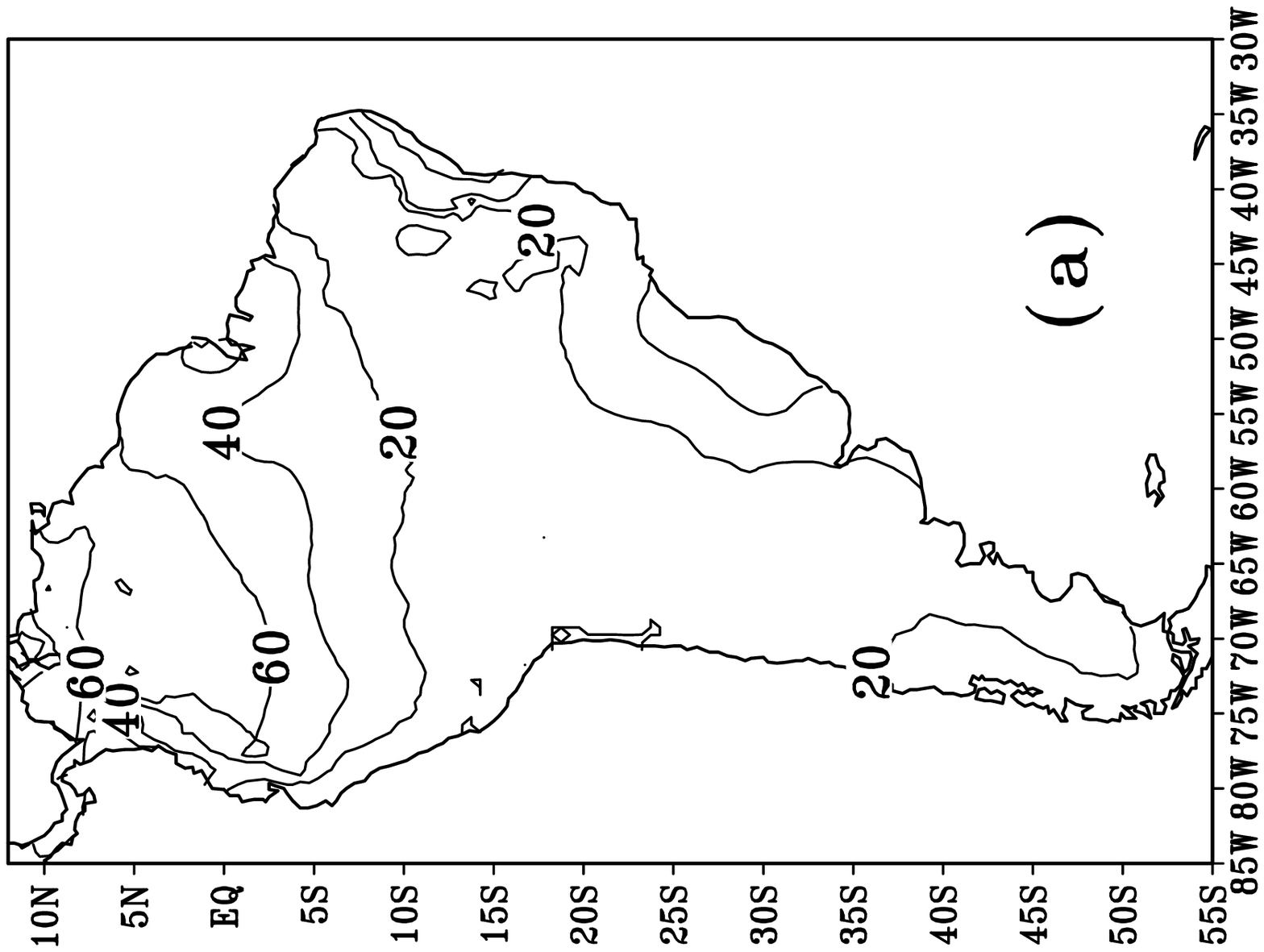}      
      \includegraphics[angle=270, width=60mm]{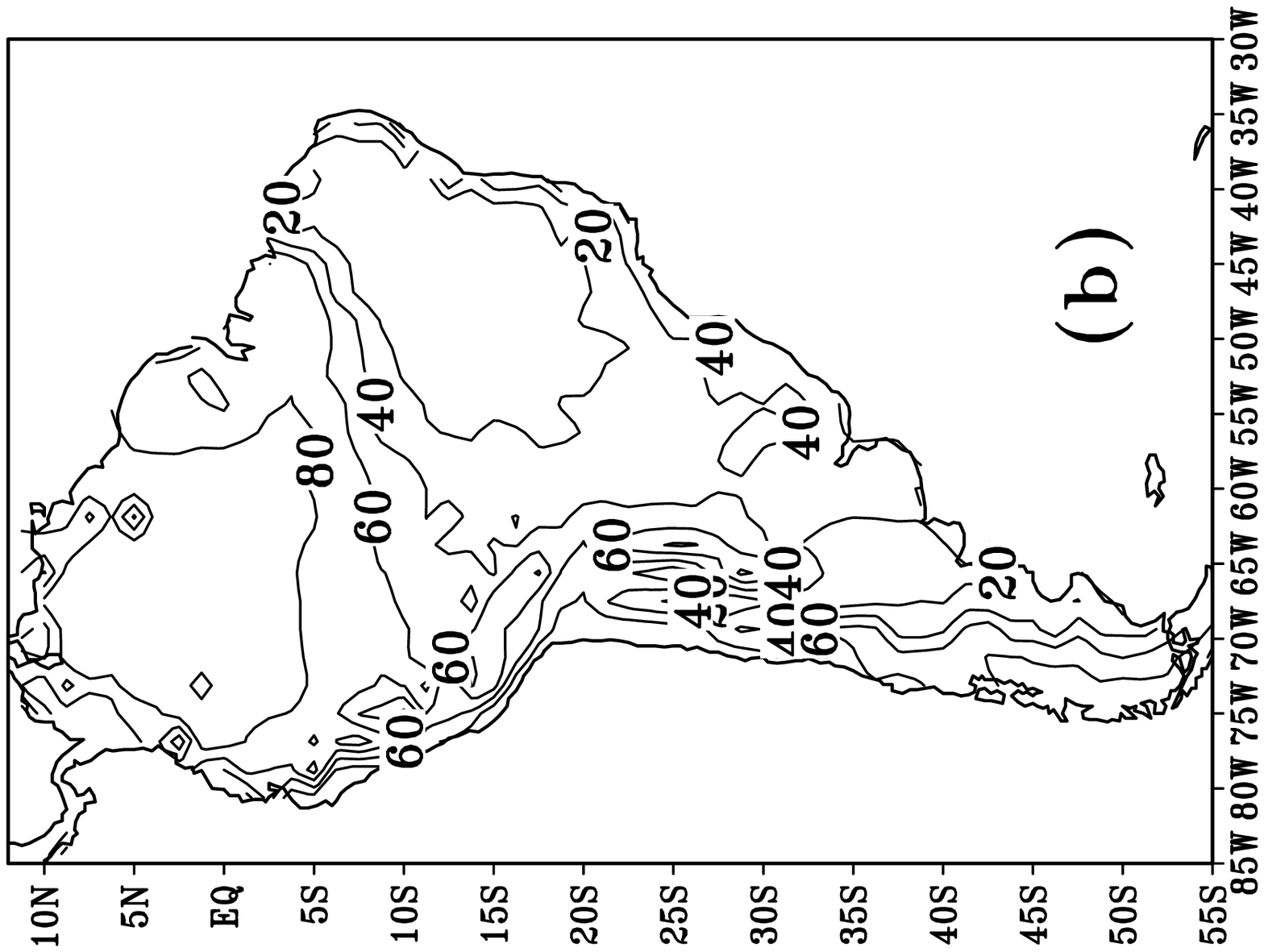}      
      }

\centerline{
      \includegraphics[angle=270, width=60mm]{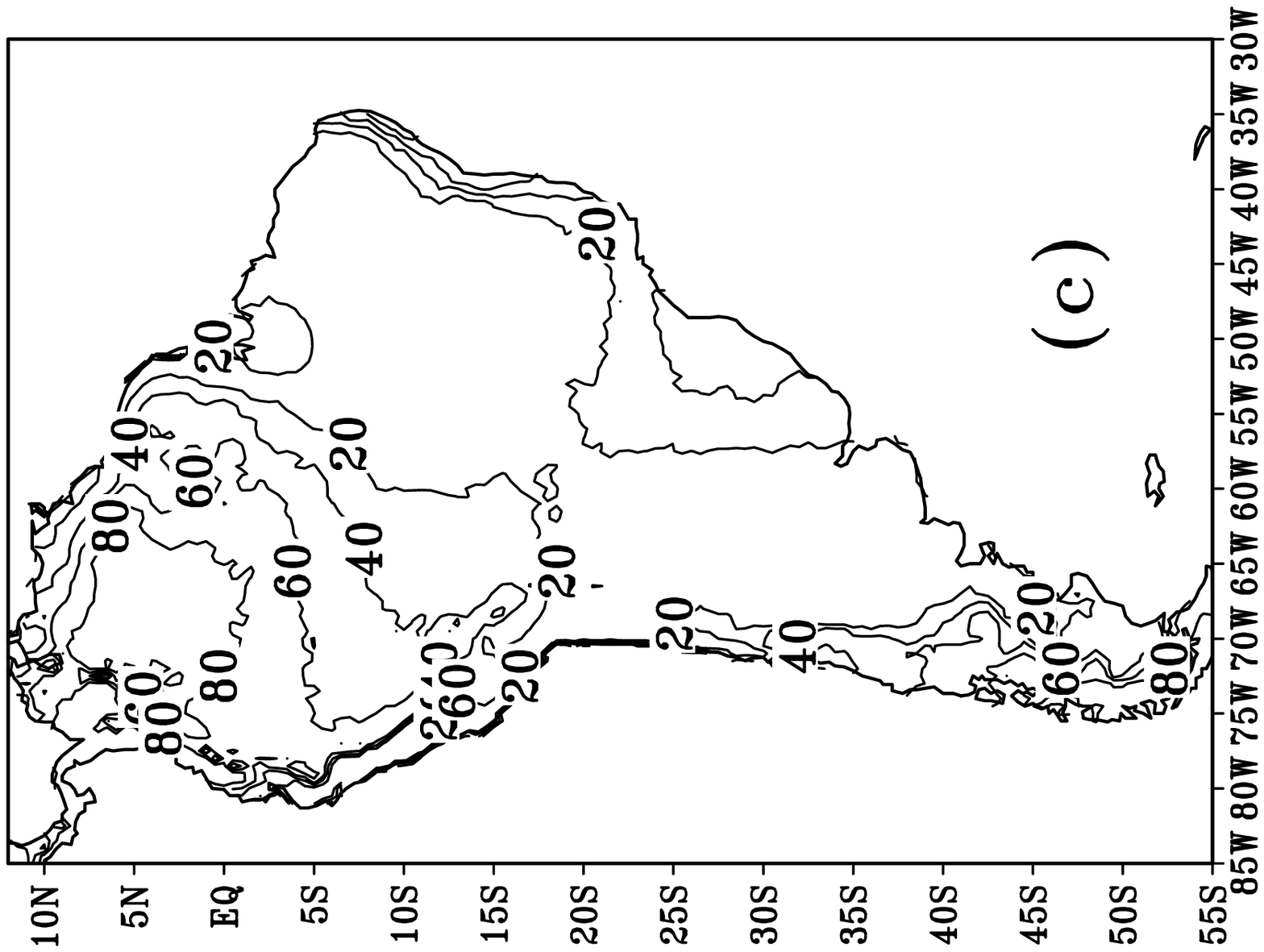}
      \includegraphics[angle=270, width=60mm]{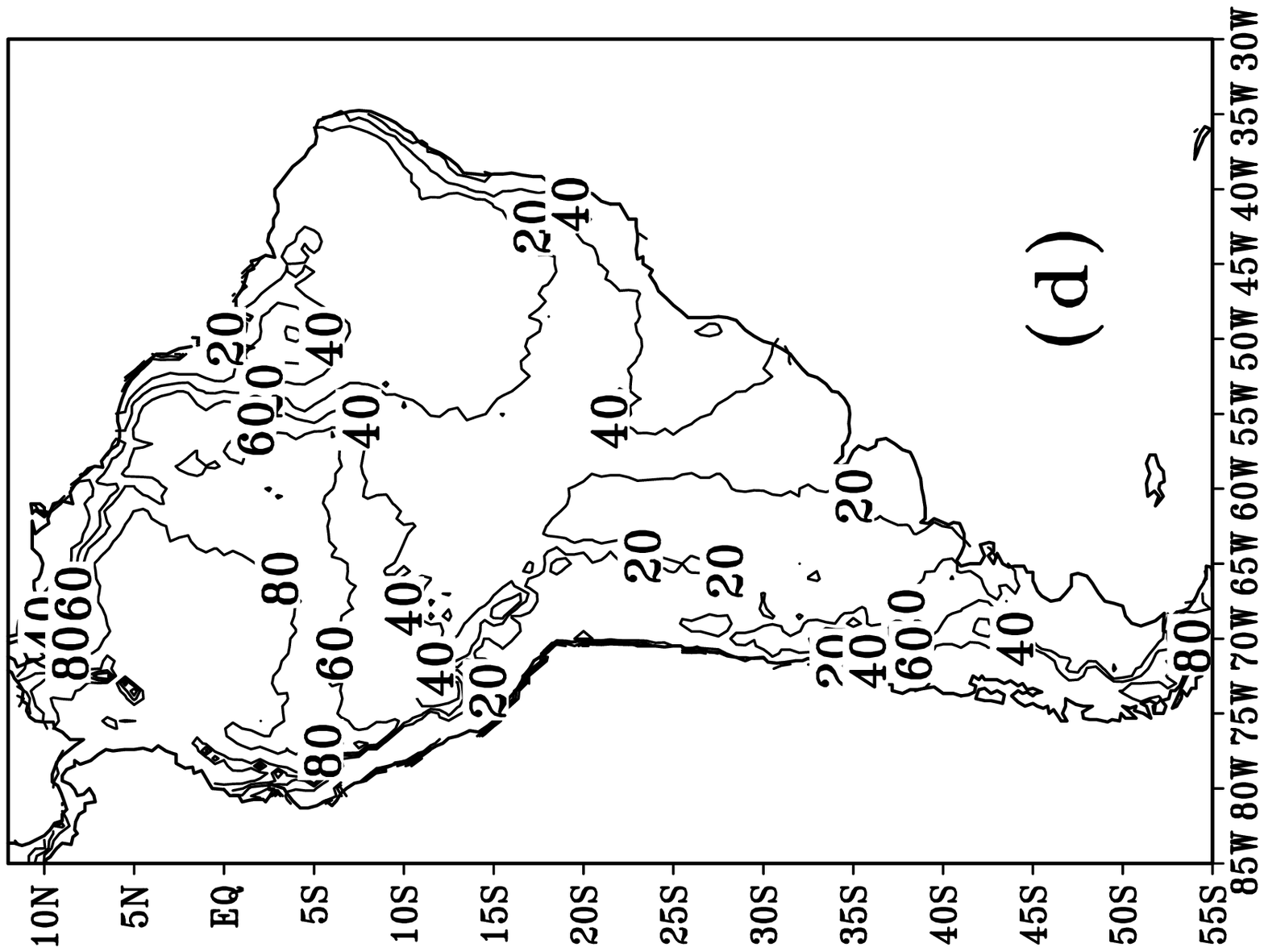}      
      }
 
\bigskip     
\caption[1]{\label{1} Number of wet days during austral winter (JJA) averaged 
over 1980-1983 years: (a) CRU, (b) HadAM3P, (c) 
Had-Eta CCS, (d) R2-Eta CCS. 
      }  
\end{figure}
\clearpage

\begin{figure}[p]
\centerline{      
      \includegraphics[angle=270, width=53mm]{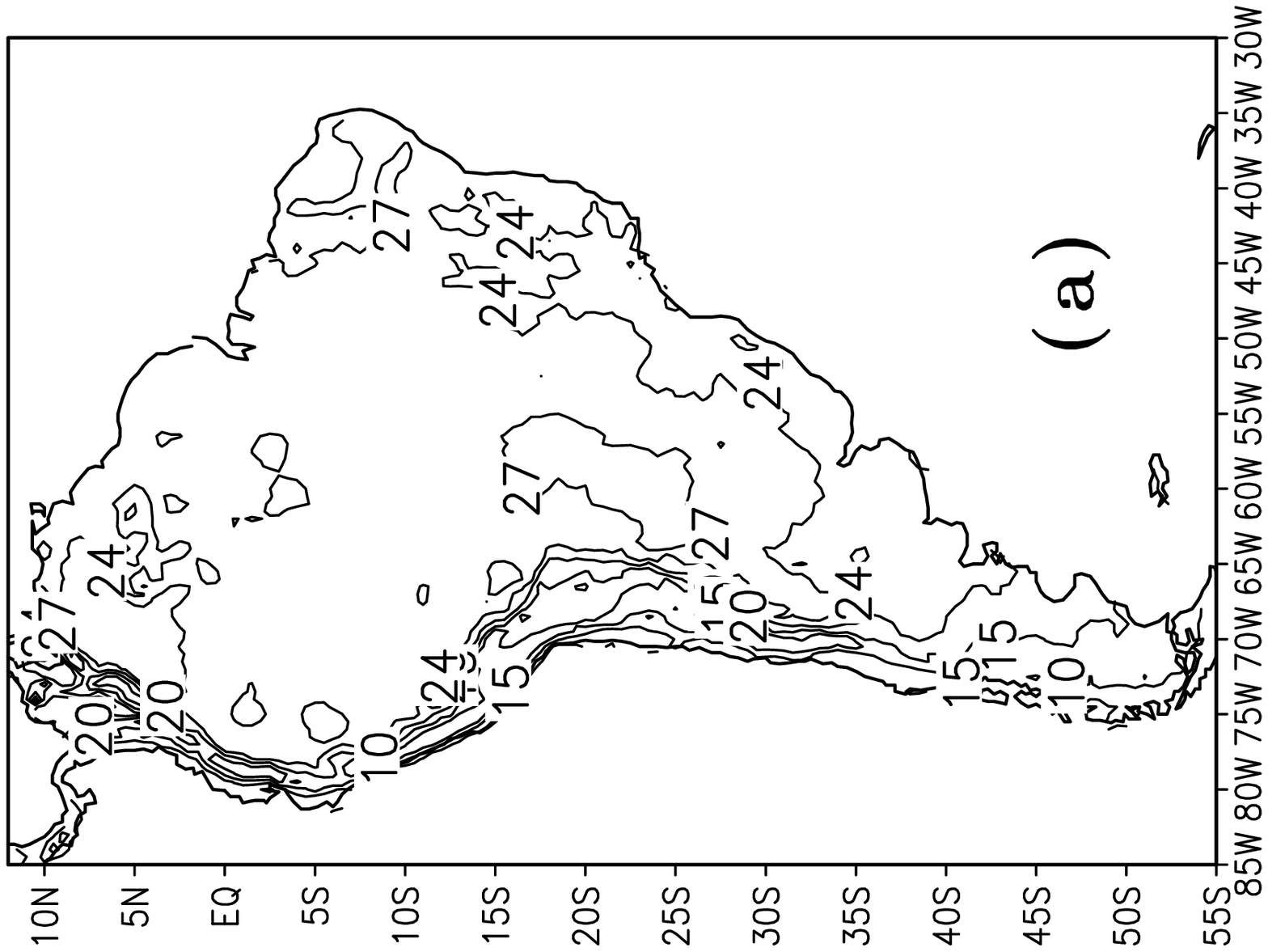}
      \includegraphics[angle=270, width=53mm]{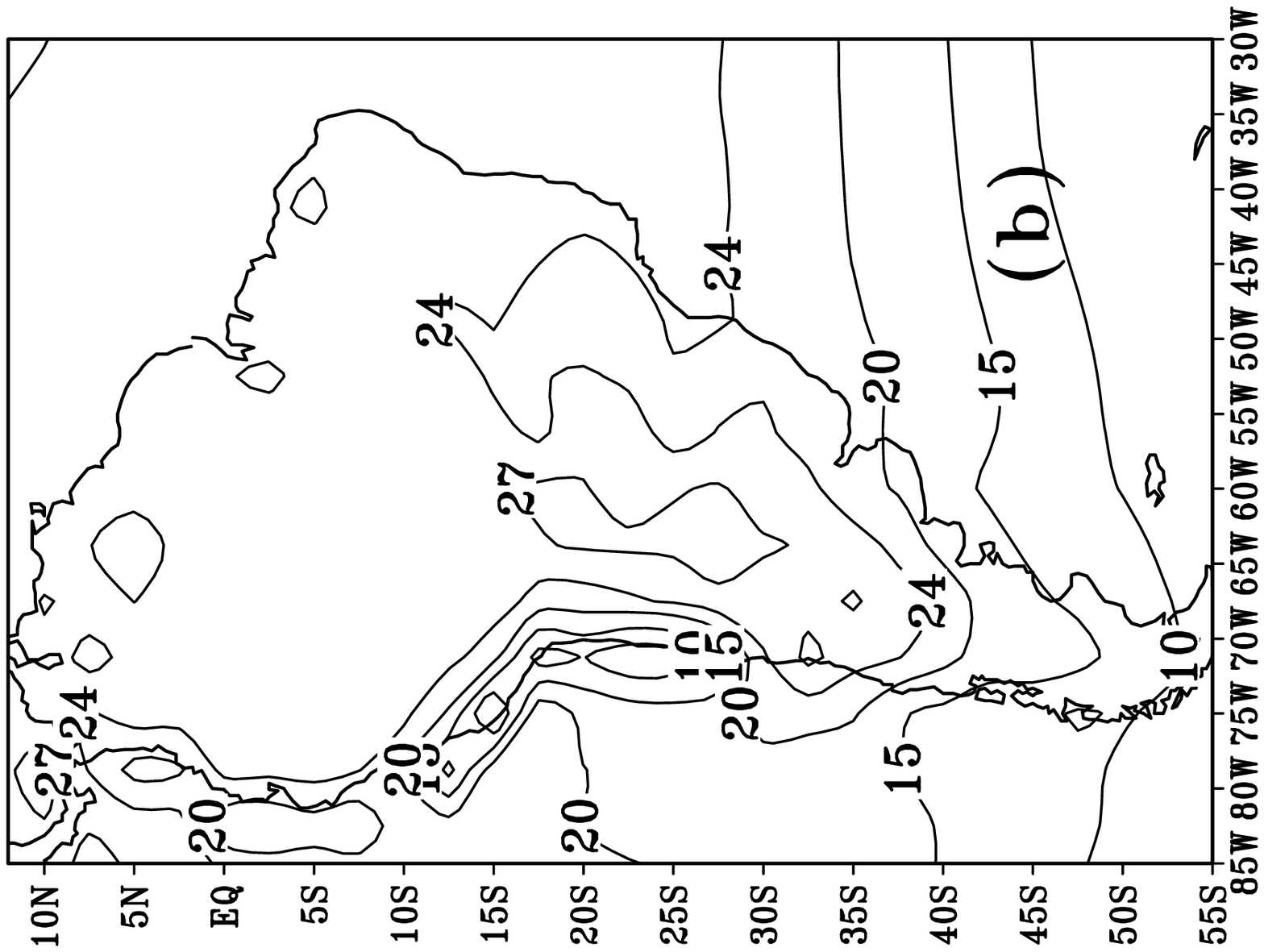} 
      \includegraphics[angle=270, width=53mm]{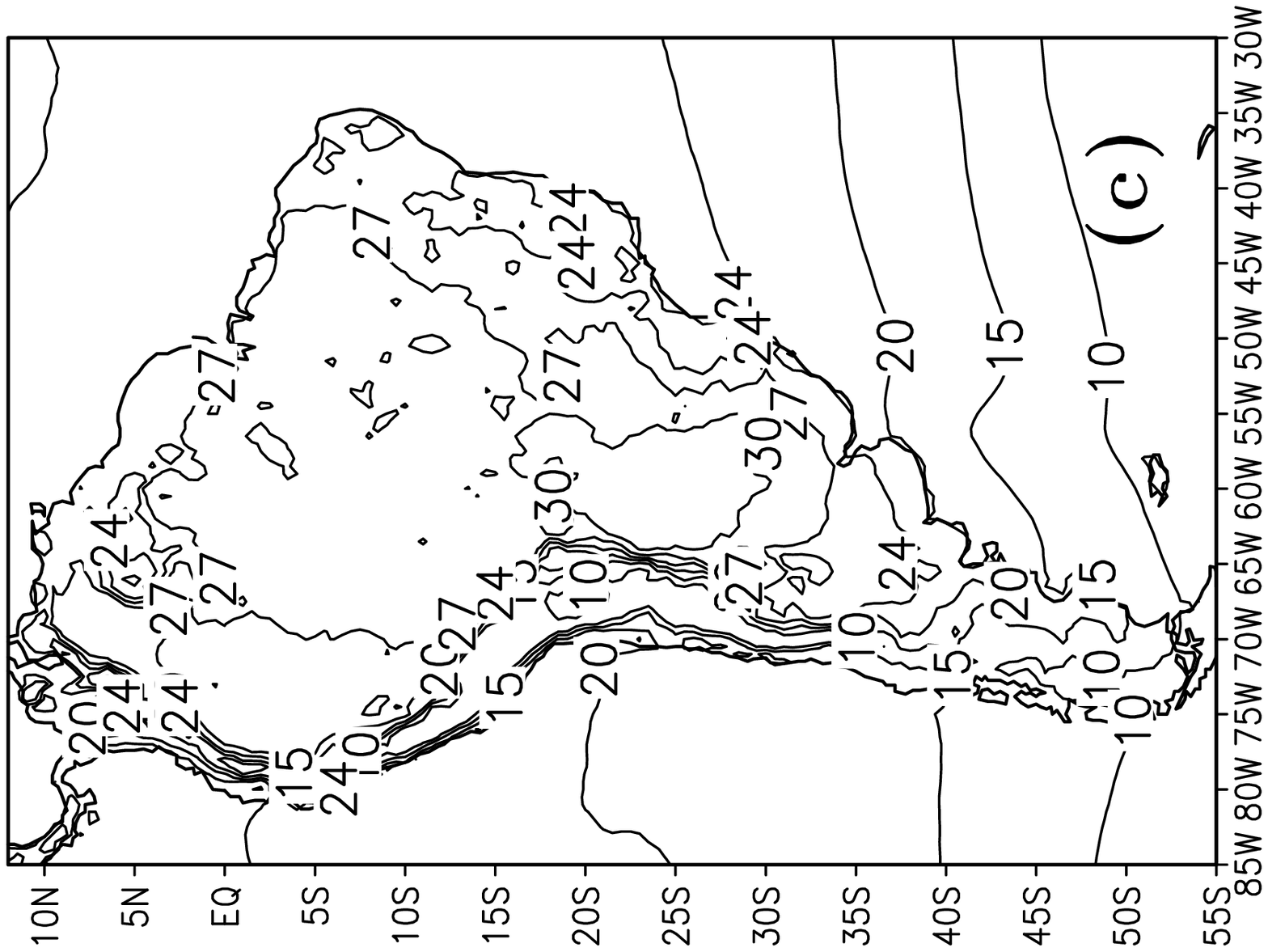}           
      }

\bigskip     
\caption[1]{\label{1} DJF mean near surface air temperature ($^{\circ}$C) averaged 
over 1980-1983 years: (a) CRU, (b) HadAM3P, (c) 
Had-Eta CCS. 
      }  
\end{figure}
\clearpage

\begin{figure}[p]
\centerline{      
      \includegraphics[angle=270, width=53mm]{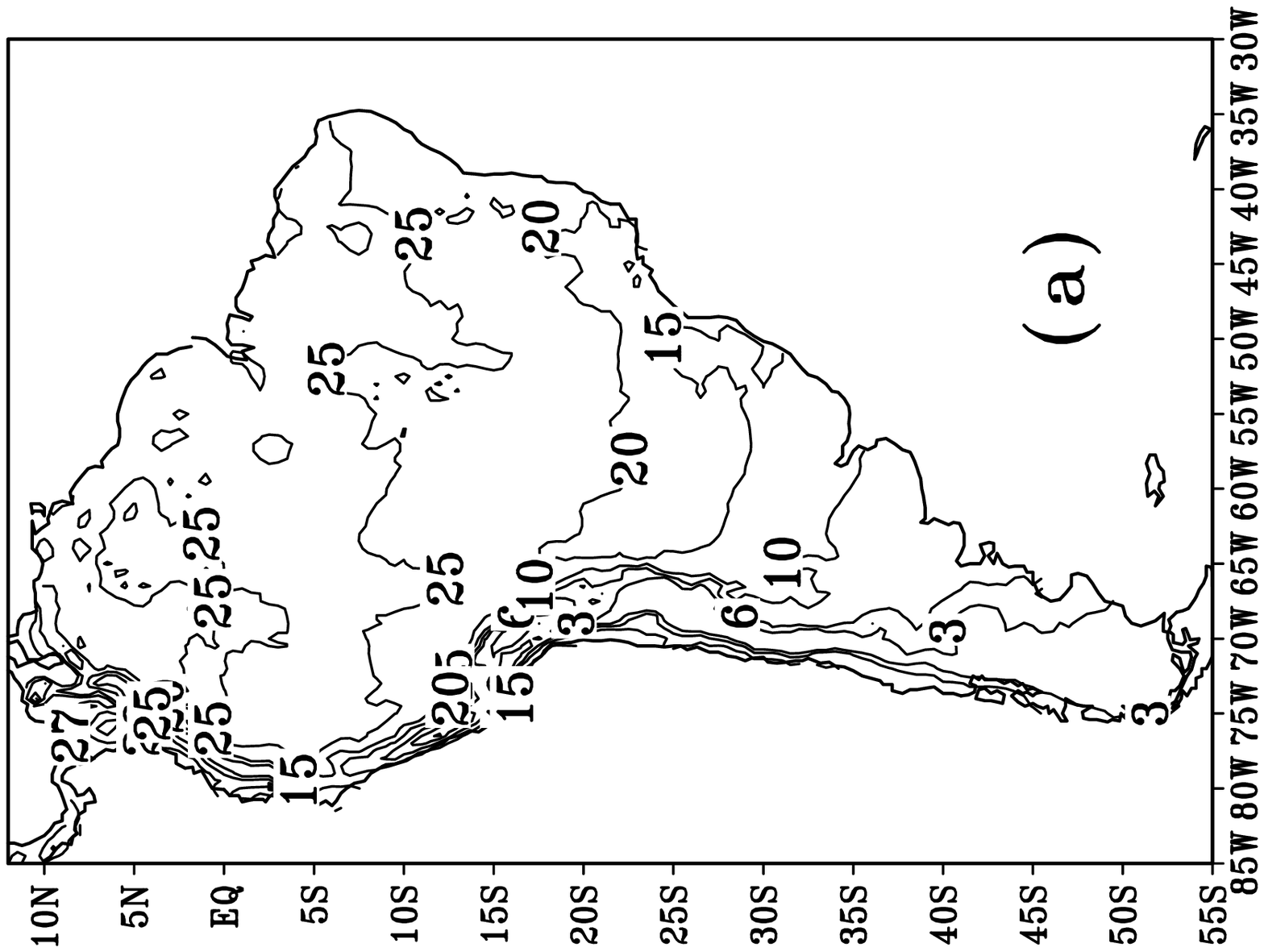}
      \includegraphics[angle=270, width=53mm]{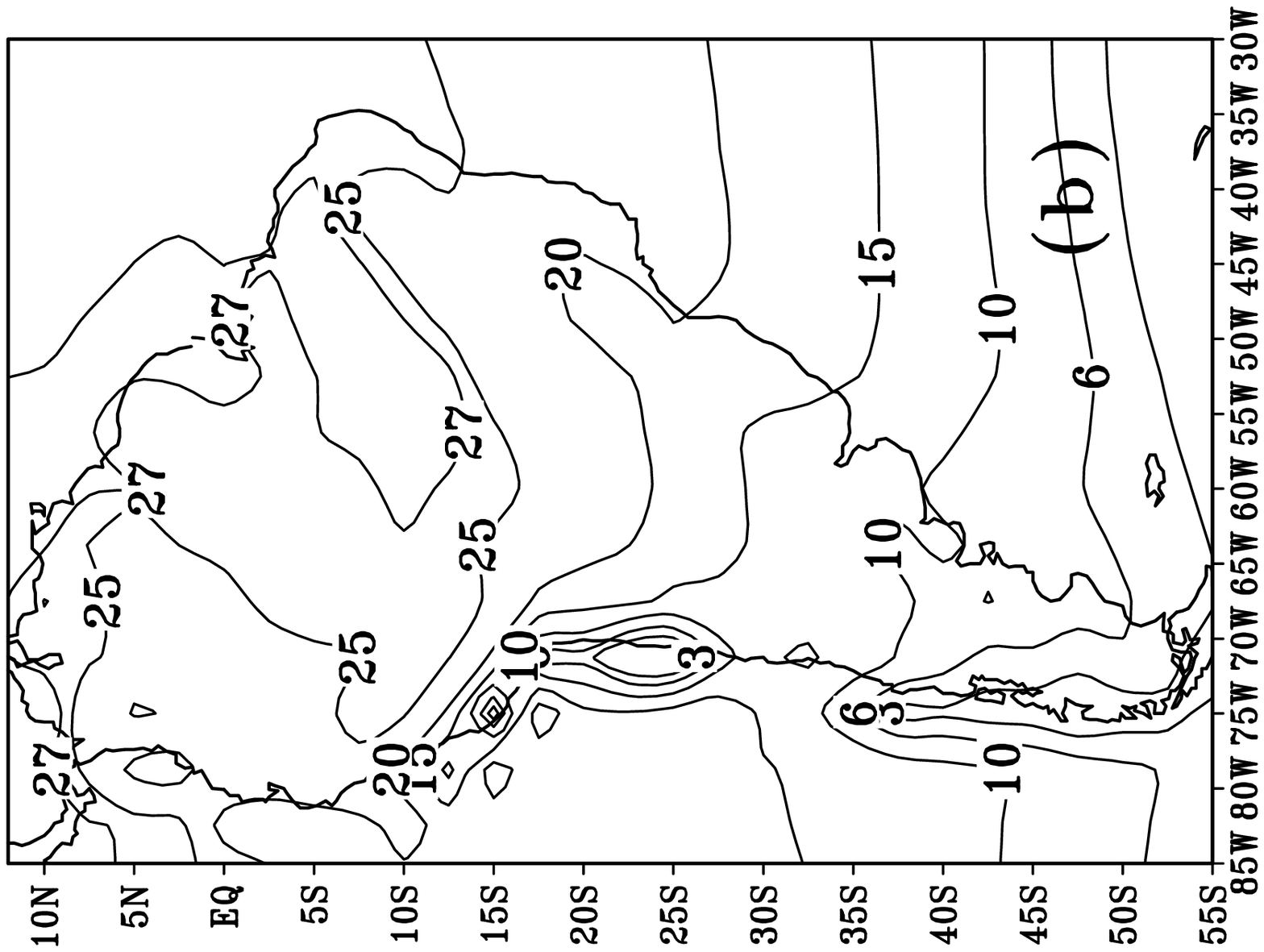} 
      \includegraphics[angle=270, width=53mm]{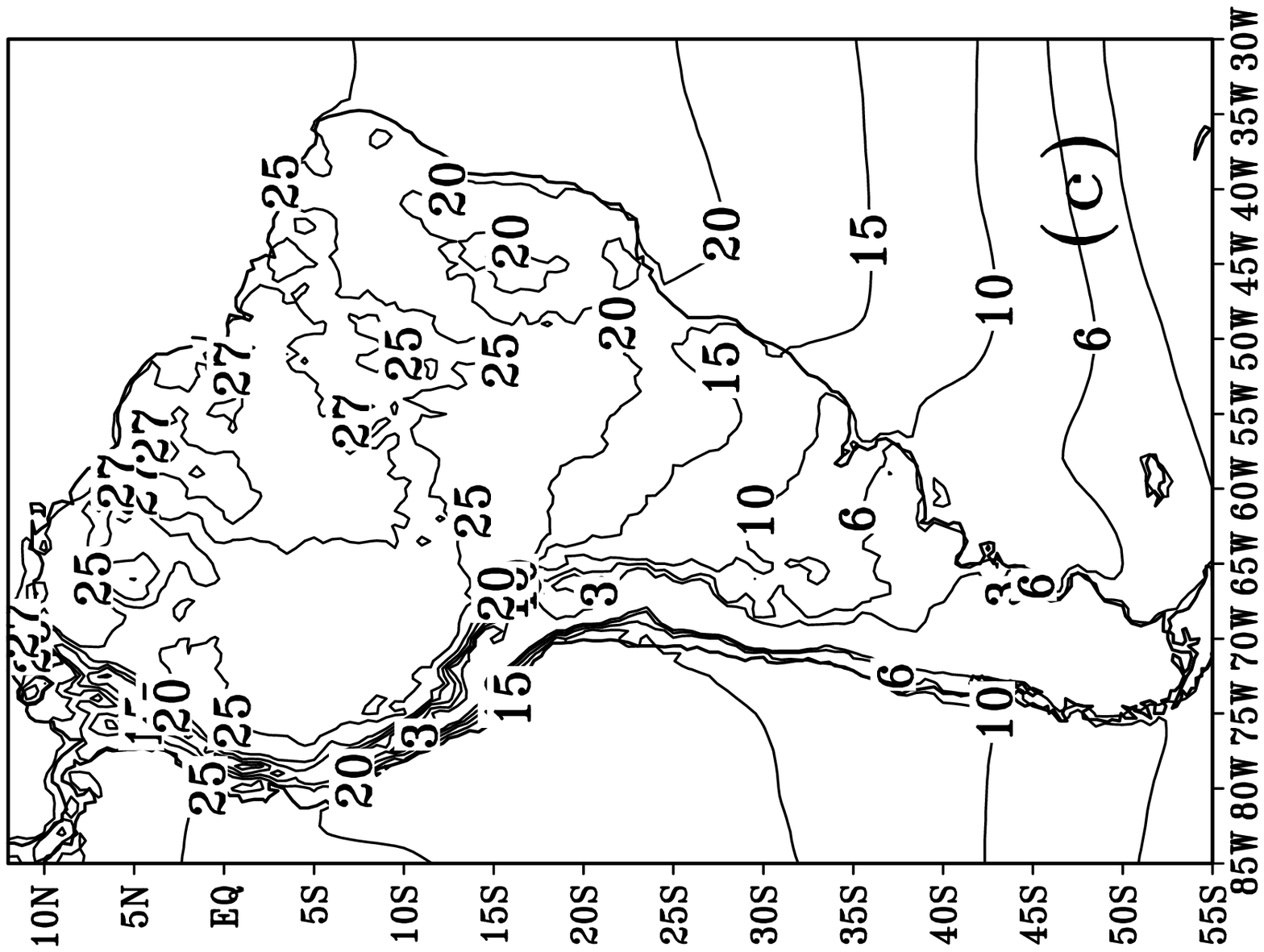}           
      }

\bigskip     
\caption[1]{\label{1} JJA mean near surface air temperature ($^{\circ}$C) averaged 
over 1980-1983 years: (a) CRU, (b) HadAM3P, (c) 
Had-Eta CCS. 
      }  
\end{figure}
\clearpage

\begin{figure}[p]
\centerline{      
      \includegraphics[angle=270, width=84mm]{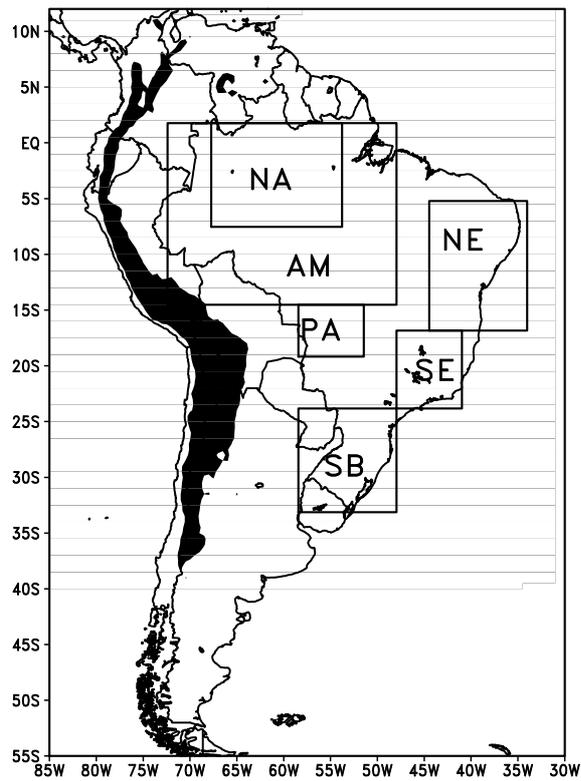}        
      }

\bigskip     
\caption[1]{\label{1} The 6 regions selected for the analysis: 
Amazon (AM),North Amazon (NA), North East Brazil (NE), South Brazil (SB), 
South East Brazil (SE), Pantanal (PA).
      }  
\end{figure}

\clearpage

\begin{figure}[p]
\centerline{      
      \includegraphics[angle=0, width=155mm]{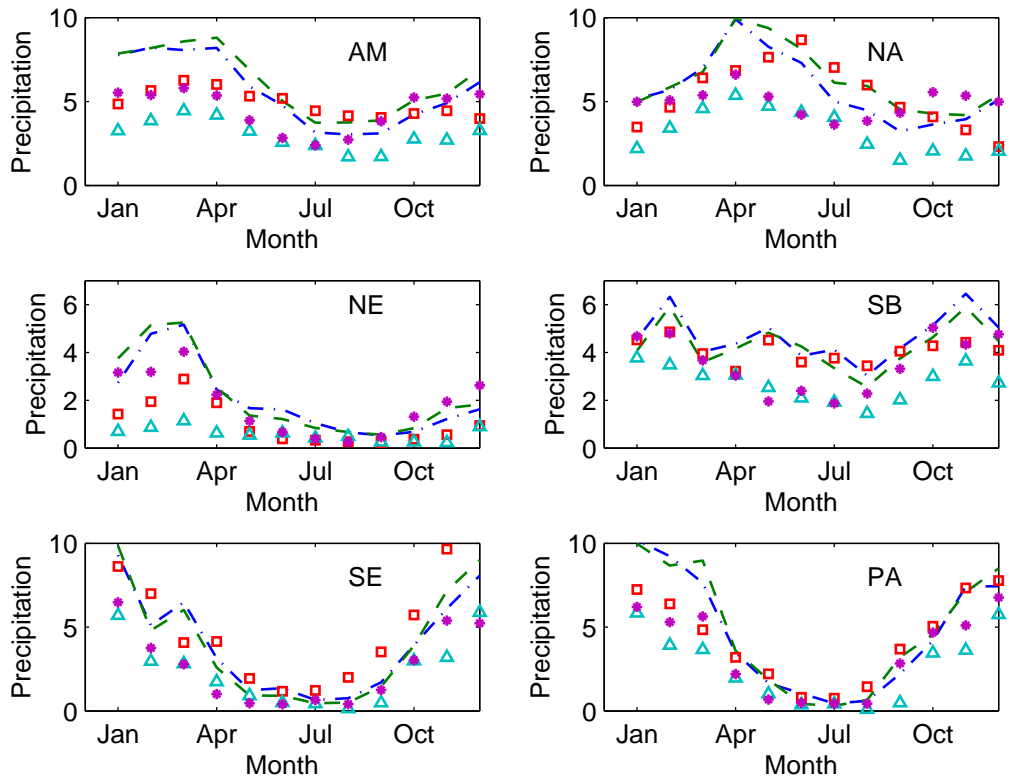}        
      }

\bigskip     
\caption[1]{\label{1} Annual cycle of monthly mean daily precipitation (mm\,d$^{-1}$)
averaged over 1980-1983 years and 6 regions (Amazon (AM), North Amazon
(NA), North East Brazil (NE), South Brazil (SB), South East Brazil (SE), 
Pantanal (PA)) from CRU (dashed), GPCP (dot-dashed), R2-Eta CCS (square),  
HadAM3P (filled circle), Had-Eta CCS (triangle).   
      }  
\end{figure}
\clearpage

\begin{figure}[p]
\centerline{      
      \includegraphics[angle=0, width=155mm]{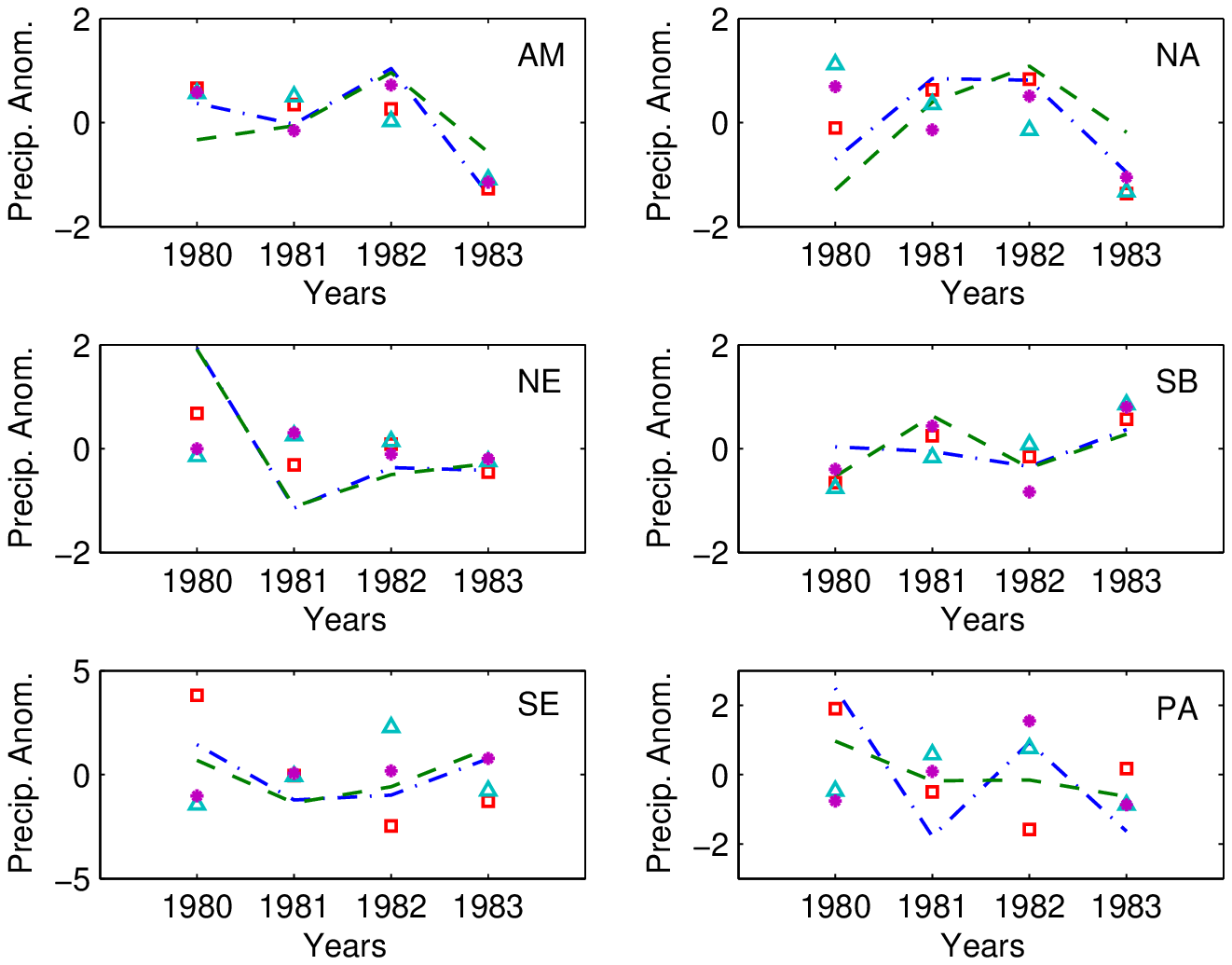}        
      }

\bigskip     
\caption[1]{\label{1} DJF mean precipitation anomaly (mm\,d$^{-1}$)
averaged over 1980-1983 years and 6 regions (Amazon (AM), North Amazon
(NA), North East Brazil(NE), South Brazil (SB), South East Brazil(SE), 
Pantanal (PA)) from CRU (dashed), GPCP (dot-dashed), R2-Eta CCS (square), 
HadAM3P (filled circle), Had-Eta CCS (triangle).    
      }  
\end{figure}
\clearpage

\begin{figure}[p]
\centerline{      
      \includegraphics[angle=0, width=155mm]{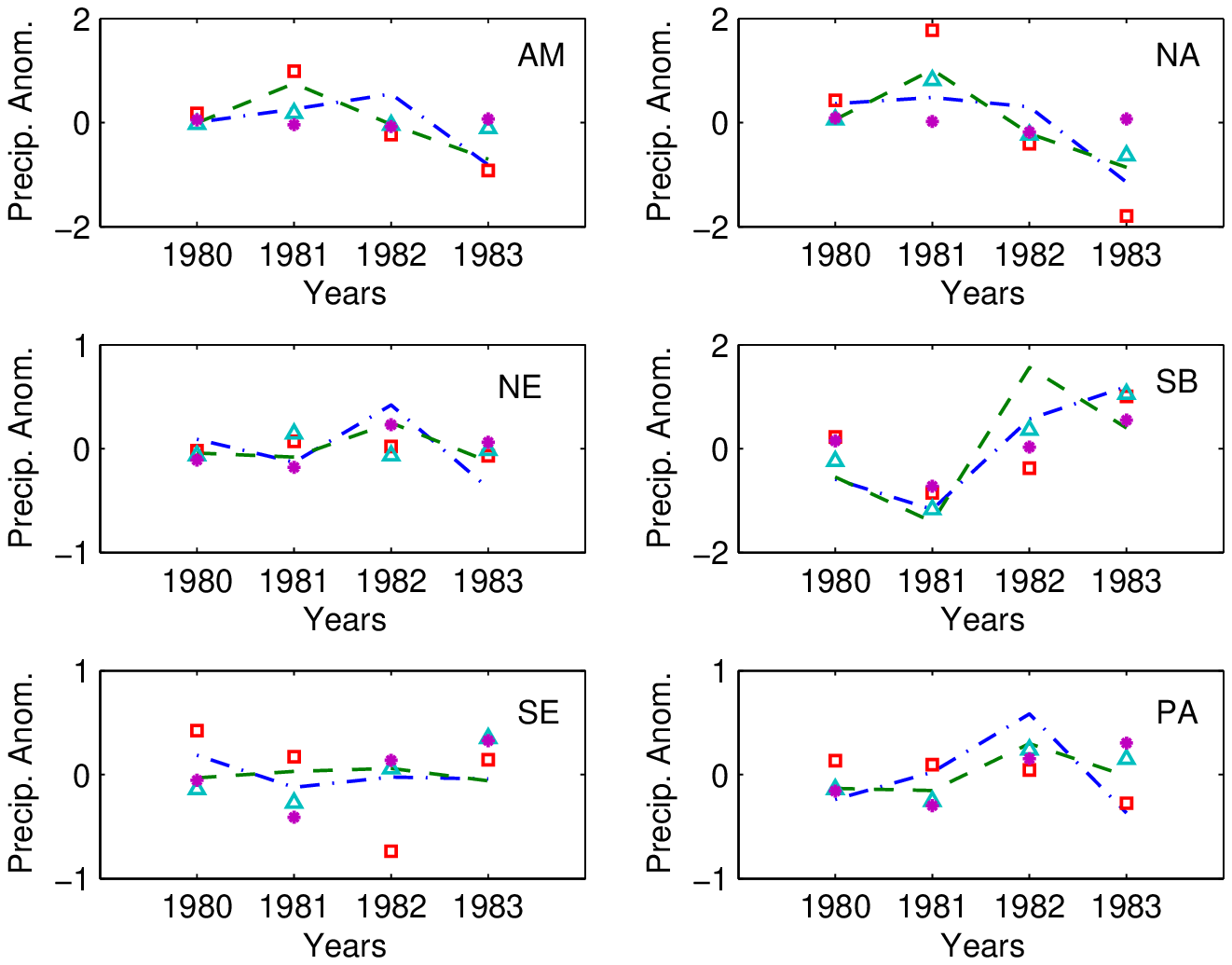}        
      }

\bigskip     
\caption[1]{\label{1} JJA mean precipitation anomaly (mm\,d$^{-1}$)
averaged over 1980-1983 years and 6 regions (Amazon (AM), North Amazon
(NA), North East Brazil (NE), South Brazil (SB), South East Brazil(SE), 
Pantanal (PA)) from CRU (dashed), GPCP (dot-dashed), R2-Eta CCS (square), 
HadAM3P (filled circle), Had-Eta CCS (triangle).   
      }  
\end{figure}

\end{document}